\documentclass[nofootinbib,preprintnumbers,prd,superscriptaddress,twocolumn,showpacs]{revtex4-1}

\usepackage{amsmath}
\usepackage{amsfonts}
\usepackage{amssymb}
\usepackage{graphicx}
\usepackage[titletoc]{appendix}
\usepackage{color}
\usepackage{hyperref}
\usepackage{cleveref}
\usepackage[rightcaption]{sidecap}
\usepackage{subfigure}
\usepackage{comment}
\usepackage{soul}
\usepackage{cancel}
\usepackage{dcolumn}

\usepackage{array}
\usepackage{ctable}
\usepackage{multirow}
\usepackage{siunitx}
\usepackage{longtable}
\usepackage{tabularx}
\usepackage{booktabs}
\usepackage{float}

\graphicspath{{Graphics/}}

\def\be{\begin{equation}}
\def\ee{\end{equation}}
\def\bea{\begin{eqnarray}}
\def\eea{\end{eqnarray}}

\def\nat{Nature}
\def\prl{Phys. Rev. Lett.}

\def\prd{Phys. Rev. D}
\def\mnras{MNRAS}
\def\aj{AJ}
\def\apj{ApJ}
\def\apjl{ApJ Lett.}
\def\apjs{ApJ Suppl. Ser.}

\def\aap{A\&A}

\def\physrep{Phys. Rep.}
\def\jcap{JCAP}

\definecolor{vividviolet}{rgb}{0.62, 0.0, 1.0}
\definecolor{amaranth}{rgb}{0.9, 0.17, 0.31}
\definecolor{palatinateblue}{rgb}{0.15, 0.23, 0.89}
\definecolor{brightpink}{rgb}{1.0, 0.0, 0.5}
\definecolor{cornflowerblue}{rgb}{0.39, 0.58, 0.93}
\definecolor{deepcarminepink}{rgb}{0.94, 0.19, 0.22}
\definecolor{radicalred}{rgb}{1.0, 0.21, 0.37}

\hypersetup{ linktoc=all,
    colorlinks, linkcolor={palatinateblue},
    citecolor={brightpink}, urlcolor={amaranth}
}

\begin{document}

\title{Cosmological transition epoch from  gamma-ray burst correlations}

\author{Anna Chiara Alfano}
\email{a.alfano@ssmeridionale.it}
\affiliation{Scuola Superiore Meridionale, Largo S. Marcellino 10, 80138 Napoli, Italy.}
\affiliation{Istituto Nazionale di Fisica Nucleare (INFN), Sezione di Napoli Complesso Universitario Monte S. Angelo, Via Cinthia 9 Edificio G, 80138 Napoli, Italy.}

\author{Salvatore Capozziello}
\email{capozziello@na.infn.it}
\affiliation{Scuola Superiore Meridionale, Largo S. Marcellino 10, 80138 Napoli, Italy.}
\affiliation{Istituto Nazionale di Fisica Nucleare (INFN), Sezione di Napoli Complesso Universitario Monte S. Angelo, Via Cinthia 9 Edificio G, 80138 Napoli, Italy.}
\affiliation{Dipartimento di Fisica "E. Pancini", Universit\`a di Napoli ``Federico II'', Complesso Universitario Monte S. Angelo, Via Cinthia 9 Edificio G, 80126 Napoli, Italy.}

\author{Orlando Luongo}
\email{orlando.luongo@unicam.it}
\affiliation{Universit\`a di Camerino, Divisione di Fisica, Via Madonna delle carceri 9, 62032 Camerino, Italy.}
\affiliation{SUNY Polytechnic Institute, 13502 Utica, New York, USA.}
\affiliation{INFN, Sezione di Perugia, Perugia, 06123, Italy.}
\affiliation{INAF - Osservatorio Astronomico di Brera, Milano, Italy.}
\affiliation{Al-Farabi Kazakh National University, Al-Farabi av. 71, 050040 Almaty, Kazakhstan.}

\author{Marco Muccino}
\email{marco.muccino@lnf.infn.it}
\affiliation{Universit\`a di Camerino, Divisione di Fisica, Via Madonna delle carceri 9, 62032 Camerino, Italy.}
\affiliation{Al-Farabi Kazakh National University, Al-Farabi av. 71, 050040 Almaty, Kazakhstan.}
\affiliation{ICRANet, P.zza della Repubblica 10, 65122 Pescara, Italy.}

\begin{abstract}
The redshift $z_t$ and the jerk parameter $j_t$ of the transition epoch can be constrained by using two model-independent approaches involving the direct expansion of the Hubble rate and the expansion of the deceleration parameter around $z=z_t$. To extend our analysis to high-redshifts, we employ the  \emph{Amati}, \emph{Combo}, \emph{Yonetoku} and \emph{Dainotti} gamma-ray burst correlations. The \textit{circularity problem} is prevented by calibrating these correlations through the B\'ezier interpolation of the updated observational Hubble data. Each gamma-ray burst data set is jointly fit with type Ia supernovae and baryonic acoustic oscillations through a Monte Carlo analysis, based on the Metropolis-Hastings algorithm, to obtain $z_t$, $j_t$ and the correlation parameters. The overall results are compatible with the concordance model with some exceptions. We also focus on the behaviors of the dark energy, verifying its compatibility with a cosmological constant, and the matter density $\Omega_m$ and compare them with the expectations of the concordance paradigm.
\end{abstract}

\pacs{98.70.Rz, 98.80.-k, 98.80.Es, 98.62.Py}


\maketitle

\section{Introduction}
\label{introduction}

The cosmic speed up is a widely-consolidated evidence, currently supported by several observations \citep{2001ApJ...560...49R, 2008Natur.451..541G, 2013PhR...530...87W} and firstly certified by type Ia supernovae (SNe Ia)  \citep{riess, Perlmutter}. The on-set of this accelerated phase occured as dark energy started to dominate over matter \textit{i.e}, in a \emph{transition epoch} marked by a transition redshift, $z_t$.

The simplest model that explains this feature, the $\Lambda$CDM paradigm, is based on the dark energy in the form of a cosmological constant, $\Lambda$ \citep{2001LRR.....4....1C, 2000astro.ph..5265W, 2003RvMP...75..559P}, and a cold dark matter contribution.
Thus, while matter decelerates the expansion of the universe, the cosmological constant accelerates it acting as a repulsive gravity \citep{2003PhR...380..235P}.

However, the cosmological constant hypothesis purported by the $\Lambda$CDM model suffers from a severe \textit{fine-tuning} problem --- if the physical interpretation of $\Lambda$ is attributed to the vacuum energy density, then a $121$ order of magnitudes discrepancy exists between predictions and observations \citep{copeland, capozziello2013} --- and a \textit{coincidence} problem --- the current densities of matter and $\Lambda$ are strangely compatible.

To address these problems, extensions of the cosmological constant paradigm have been proposed. The simplest ones are the so-called $\omega$CDM model, in which the equation of state is constant \citep{2002CQGra..19.3435S,2013CQGra..30u4003T}, and the Chevallier-Polarksy-Linder (CPL) parametrization \citep{2001IJMPD..10..213C, 2003PhRvL..90i1301L}, where the equation of state is written as a function of the scale factor $a(t)$.

Alternatively, model-independent approaches have been proposed to estabilish whether dark energy behaves as a cosmological constant or evolves with time \citep{2003PhRvD..67f3521C, 2007MNRAS.380.1573S, 2013PhRvD..88f3521N, 2014ApJ...793L..40S}. To understand this, it is essential to investigate the transition epoch resorting the cosmographic approach \citep{Harrison, Visser, 2016IJGMM..1330002D, salzano}, based on Taylor series of the Hubble rate through the cosmographic parameters, whose number depends upon the selected order of the expansion \citep{2007CQGra..24.5985C,2008PhRvD..78f3501C, 2012PhRvD..86l3516A, 2014PhRvD..89j3506G}.
When applied to the transition epoch, the key cosmographic quantities involved in the expansion are the jerk parameter at the transition $j(z_t)=j_t$ and the redshift at the transition $z_t$, whereas the deceleration parameter, by definition vanishes, \emph{i.e.}, $q(z_t)=q_t=0$.

Here, we work with the direct expansion of the Hubble rate (DHE) and the direct deceleration parameter expansion (DDPE) around $z\simeq z_t$ \citep{capozziellodunsbyluongo}. Then, we get constraints on the transition redshift and cosmographic parameters at the transition by adopting standard candles, SNe Ia, standard rulers, baryonic acoustic oscillations (BAO), and gamma-ray bursts (GRB), a class of distance indicators enabling the investigation of the universe at high-redshift \citep{ghirlanda1, firmani, luongomuccino2021, ester1,ester2, ester3, giada1,giada2}.
However, GRB need to be calibrated in a model-independent way to avoid the well-known \textit{circularity} problem \citep{luongomuccino2021} and, to this aim, we here resort to the B\'ezier interpolation \citep{bezier, 2021MNRAS.503.4581L, 2023MNRAS.518.2247L, 2023arXiv231105324A} to model the Hubble rate and, thus, the cosmological distances. We apply this technique to four GRB correlations, namely Amati, Combo, Yonetoku and Dainotti's correlations \citep{amatiref, comboref, Yonetoku, 2022MNRAS.512..439C} using the 33 updated Hubble rate data \citep{kumaretal, 2023ApJS..265...48J}.

We perform a Monte Carlo -- Markov Chain (MCMC) analyses to constrain the parameters of DHE and DDPE models and of the four GRB correlations. Then, we apply our results to study the dark energy behavior, constrain the matter density $\Omega_m$, and compare them with the $\Lambda$CDM paradigm.

The paper is structured as follows.
In Sec.~\ref{sec2}, we describe the GRB correlations we are going to utilize and, in Sec.~\ref{sec3}, we describe how to calibrate them in a model-independent way through the \emph{B\'ezier interpolation} technique.
In Sec.~\ref{sec4}, we illustrate the theoretical methods we adopt to constrain the transition epoch.
In Sec.~\ref{sec5}, we describe our MCMC calculations and summarize the results that are applied in Sec.~\ref{sec6} to analyze the behavior of the dark energy at $z_t$ and to compute the values of the matter density $\Omega_m$.
Finally, in Sec.~\ref{sec7}, we present our conclusions.

\section{GRB correlations}\label{sec2}

GRB are high-energy astrophysical sources of $\gamma$-rays, essential to probe the universe up to $z\sim 9$ \citep{Salvaterra2009,Cucchiara2011}.
To this aim, several linear correlations involving photometric and spectroscopic properties have been proposed as tools to standardize GRB and render them distance indicators \citep{amatiref, comboref, Yonetoku, 2022MNRAS.512..439C,ghirlanda}.

We use two GRB correlations based on the prompt emission, and two using both prompt and X-ray afterglow emissions.

\subsection{Prompt emission correlations}

The prompt emission of a GRB is generally a multi-peaked and highly-variable light curve observed at $\gamma$-ray energies, with total observed duration $t_{90}$, evaluated as the time interval over which from $90$\% of the total background-subtracted counts are observed \citep{2021Galax...9...77L}.
The corresponding $\gamma$-ray energy spectrum, integrated over the $t_{90}$ or in sub-intervals, is generally best-fit by a smoothly-joined double power-law model, peaking at a specific observed energy $E_p^o$ \cite{Band2003}.

Below we provide the details of two well-established GRB prompt emission correlations.
\begin{itemize}
\item[-] \textbf{Amati or $E_p$--$E_{iso}$ correlation.}
This correlation is expressed as \citep{amatiref}
\begin{equation}
\label{Amati_correlation}
    \log\left(\frac{E_p}{{\rm keV}}\right)=a\left[\log\left(\frac{E_{iso}}{{\rm erg}}\right)-52\right]+b\,,
\end{equation}
where, for each GRB, we have the rest-frame peak energy $E_p=(1+z)E_p^o$, the burst isotropic radiated energy
\begin{equation}\label{Eiso}
    E_{iso} = 4\pi d_L^2(z,p)S_b(1+z)^{-1}\,,
\end{equation}
the observed bolometric fluence $S_b$, computed in the rest-frame $1$--$10^4$~keV band, and $d_L$ is the luminosity distance.
\item[-] \textbf{Yonetoku or $L_p$--$E_p$ correlation.}
The expression of this correlation is \citep{Yonetoku}
\begin{equation}
\label{Yonetoku_correlation}
    \log\left(\frac{L_p}{{\rm erg/s}}\right) = a\log\left(\frac{E_p}{{\rm keV}}\right)+b+52,
\end{equation}
where $E_p$ now correlates with the peak luminosity
\begin{equation}\label{Lp}
    L_p = 4\pi d_L(z, p)^2 F_p\,,
\end{equation}
related to the peak flux $F_p$, computed in the most intense $1$~s time interval of the burst light curve and in the rest-frame $30$--$10^4$~keV energy band \citep{Yonetoku}.
\end{itemize}

Both correlations have slope $a$, intercept $b$, and intrinsic extrascatter $\sigma$. Moreover, in Eqs.~\eqref{Eiso} and \eqref{Lp} the luminosity distance depends upon the cosmological parameters $p$, making both correlations cosmology-dependent.

\subsection{Prompt emission and afterglow correlations}

The X-ray afterglow emission is detected in the $0.3-10$~keV band, when the prompt emission begins to decay in flux. The typical X-ray afterglow light curve shows an early decay stage, followed by a plateau and a late-time decay, whereas the corresponding spectrum is best-fit by a single power-law model with index $\beta\approx2$ \citep{2021Galax...9...77L}.

Below we show two well-known, hybrid prompt--afterglow emission GRB correlations.
\begin{itemize}
\item[-] \textbf{Combo or $L_0$--$E_p$--$T$ correlation.}
This hybrid correlation has the following shape \citep{comboref}
\begin{equation}
\label{Combo_correlation}
\log \left(\frac{L_0}{{\rm erg/s}}\right) = a \log \left(\frac{E_p}{{\rm keV}}\right) - \log \left(\frac{T}{{\rm s}}\right) + b\,,
\end{equation}
and links the prompt observable $E_p$ with the luminosity $L_0$ and the rest-frame duration $\tau$ of the plateau, and the late decay index $\alpha$ of the $0.3$--$10$~keV rest-frame luminosity.
The last two quatities are combined into $T\equiv\tau/|1+\alpha|$.
The plateau luminosity is related to the plateau flux $F_0$ through
\begin{equation}\label{L0}
    L_0=4\pi d^2_L(z, p)F_0\,.
\end{equation}
\item[-]{\bf Dainotti or $L_X$--$T_X$--$L_p$ correlation.}
This is a three-parameter fundamental plane correlation \citep{2022MNRAS.512..439C}
\begin{equation}
\label{Dainotti_correlation}
    \log\left(\frac{L_X}{{\rm  erg/s}}\right) = a\log\left(\frac{T_X}{{\rm s}}\right) + b\log\left(\frac{L_p}{{\rm erg/s}}\right)+c\,,
\end{equation}
where, in this representation, the prompt peak luminosity is related to the peak flux via
\begin{equation}\label{LpD}
    L_p = 4\pi d^2_L(z,p)F_p(1+z)^{-(1-\beta)}\,,
\end{equation}
and the X-ray rest-frame luminosity $L_X$ relates to the X-ray flux $F_X$ at the rest-frame time $T_X$ at the end of the plateau
\begin{equation}\label{Lx}
    L_X = 4\pi d^2_L(z,p)F_X(1+z)^{-(1-\beta)}\,,
\end{equation}
\end{itemize}

For the $L_0$--$E_p$--$T$ correlation $a$ is the slope and $b$ the intercept. The $L_X$--$T_X$--$L_p$ correlation has two slopes $a$ and $b$, and the intercept $c$. Both correlations have an extrascatter $\sigma$. Finally, the luminosity distance in Eqs.~\eqref{L0}, \eqref{LpD} and \eqref{Lx} depends upon the cosmological parameters $p$.

\section{Model-independent B\'ezier calibration}\label{sec3}

The four GRB correlations of Sec.~\ref{sec2} are all model-dependent, as they are related to $d_L(z,p)$, which depends upon the cosmological model parameters $p$.
Thus, unless we calibrate them, all GRB correlations depend upon the cosmological model we choose. This is what it is referred to as \textit{circularity problem}.

Unlike SNe Ia, GRB cannot be anchored to primary indicators due to the lack of very-low redshift sources.
Thus, they can be calibrated via model-independent techniques.

The B\'ezier interpolation \citep{bezier, 2020A&A...641A.174L, 2021MNRAS.503.4581L, 2023MNRAS.523.4938M, 2023arXiv231105324A} applied to the updated catalog of $33$ observational Hubble data (OHD) \citep{kumaretal, 2023ApJS..265...48J}, provides a powerful method to interpolate the Hubble rate $H(z)$ without postulating any \textit{a priori} cosmology. The corresponding B\'ezier curve of order $n$ reads
\begin{equation}
    H_n(x) = 100 \sum_{i=0}^{n} \alpha_i n!\frac{x^i}{i!}\frac{\left(1-x\right)^{n-i}}{(n-i)!}\ \left(\frac{{\rm km/s}}{{\rm Mpc}}\right)\,,
\end{equation}
which is positive-defined for $0\leq x\equiv z/z_O\leq 1$ with $z_O$ the highest redshift of the OHD catalog.

The only interpolating curve providing a non-linear, monotonic growing function is the B\'ezier curve of order $n=2$, labeled by $H_2(z)$ \citep{bezier, 2020A&A...641A.174L, 2021MNRAS.503.4581L, 2023MNRAS.523.4938M}.
Thus, assuming a spatially flat universe, the luminosity distance is given by
\begin{equation}\label{callumdist}
    d_L^{cal}(z)=c(1+z)\int^{z}_{0} \frac{dz'}{H_2(z')},
\end{equation}
being $c$ the speed of light, and can be used in Eqs.~\eqref{Eiso}, \eqref{Lp}, \eqref{L0} and \eqref{LpD}--\eqref{Lx} to calibrate Eqs.~\eqref{Amati_correlation}, \eqref{Yonetoku_correlation}, \eqref{Combo_correlation} and \eqref{Dainotti_correlation}.

\section{Theoretical expansions at the transition epoch}\label{sec4}

Now, we illustrate the expansion methods that will be used to provide constraints on $z_t$ and $j_t$ \citep{capozziellodunsbyluongo}.

\begin{enumerate}
    \item[-] \textbf{DHE method.} Here, we directly expand $H(z)$ around $z_t$
    \begin{equation}\label{hrateex}
        H \approx H_t + H'_t (z-z_t)+\frac{1}{2}H_t^{\prime\prime}(z-z_t)^2+\mathcal{O}[(z-z_t)^3]\,,
    \end{equation}
    where we defined $y^\prime=dy/dz$ and $y(z_t)=y_t$, and required that reduces to $H\equiv H_0$ at $z=0$.
    The cosmographic parameters and $H(z)$ are correlated to each other through
    \begin{equation}\label{Hdot}
        \Dot{H}=-H^2(1+q) \quad,\quad \Ddot{H}=H^3(j+3q+2),
    \end{equation}
    where $\dot{y}=dy/dt$, $q$ is the deceleration parameter and $j$ is the jerk parameter.
    Then, we use the identity $\dot z\equiv-(1+z)H$ that, combined with the first of Eqs.~\eqref{Hdot}, provides
    \begin{equation}\label{dq}
        \frac{dq}{dz} = \frac{j-2q^2-q}{1+z}\,.
    \end{equation}
   Now, combining together Eqs.~\eqref{Hdot}, we obtain
   \begin{equation}\label{Hderiv}
       H^\prime=\frac{H(1+q)}{1+z} \quad,\quad H^{\prime\prime}=\frac{H(j-q^2)}{(1+z)^2}\,.
   \end{equation}
   Recalling that $q_t=0$, at $z_t$ Eqs.~\eqref{Hderiv} become
   \begin{equation}\label{Hzt}
       H^\prime_{t}=\frac{H_t}{1+z_t} \quad \text{and} \quad H^{\prime\prime}_{t}=\frac{H_tj_t}{(1+z_t)^2}.
   \end{equation}
   We plug Eqs.~\eqref{Hzt} into Eq.~\eqref{hrateex}, then compute $H_0$ by substituting $z=0$ in Eq.~\eqref{hrateex} and replace the so-obtained $H_0$ into and Eq.~\eqref{hrateex}. Finally, defining $\xi=H/H_0$, we obtain
   \begin{equation}
       \xi(z,z_t,j_t)= 1+\frac{j_t z^2+2z\left(1+z_t-j_t z_t \right)}{2+z_{t}\left(2+j_{t}z_{t}\right)}\,.
   \end{equation}

\item[-] \textbf{DDPE method.} We directly expands $q(z)$ around $z_t$
\begin{equation}\label{qdec}
    q \approx \frac{dq}{dz}\Big|_{z_{t}}(z-z_{t})+\mathcal{O}(z-z_{t})^2\,,
\end{equation}
where $dq/dz|_{z_t}$ is given by Eq.~\eqref{dq}. From Eq.~\eqref{qdec}, we see that the deceleration parameter has a minimum and a maximum value \citep{riess2004, capozziello2019, benetti}
\begin{equation}
\begin{split}
\begin{cases}
    \displaystyle
    q_0=-\frac{j_tz_t}{1+z_t} &,\quad z=0, \\
    q_t= 0 &, \quad z=z_t\,,
    \end{cases}
\end{split}
\end{equation}
from which it is evident that $q$ switches its sign at $z_t$.
Now, to get the Hubble rate, we plug Eq.~\eqref{qdec} into the first of Eqs.~\eqref{Hderiv} and, setting the condition $H(z=0)\equiv H_0$, we get
\begin{equation}
    \xi(z,z_t,j_{t})=\exp\left(\frac{j_t}{1+z_t}z\right)(1+z)^{1-j_t}.
\end{equation}
\end{enumerate}

\section{Numerical results}\label{sec5}

In the first step of our analysis, we calibrate the GRB data sets by estimating the coefficients $\alpha_i$ (with $i=0,1,2$) of the B\'ezier interpolating function $H_2(z)$ through the maximization of the OHD log-likelihood function
\begin{equation}
\label{loglikeOHD}
    \ln\mathcal{L}_O=-\frac{1}{2}\sum^{N_O}_{i=1}\left\{\left[\frac{H_i-H_2(z_i)}{\sigma_{H_i}}\right]^2+\ln(2\pi \sigma^2_{H_i})\right\},
\end{equation}
where $N_O=33$ is the size of the OHD catalog with Hubble rate measurements $H_i$ and attached errors $\sigma_{H_i}$.

\begin{table}[H]
    \centering
    \setlength{\tabcolsep}{0.7em}
    \renewcommand{\arraystretch}{1.4}
    \begin{tabular}{ccc}
    \hline\hline
    {$\alpha_0$} & {$\alpha_1$} & {$\alpha_2$} \\ \hline
{$0.674^{+0.052(0.101)}_{-0.047(0.098)}$} & {$1.047^{+0.141(0.288)}_{-0.153(0.302)}$} & {$2.078^{+0.197(0.394)}_{-0.190(0.382)}$}  \\ \hline
    \end{tabular}
    \caption{Best-fit B\'ezier coefficients $\alpha_i$ obtained from the OHD catalog.}
    \label{tabBezier}
\end{table}

We perform an MCMC fit, based on the Metropolis-Hastings algorithm \citep{1953JChPh..21.1087M,1970Bimka..57...97H}, with a total of $20000$ steps and parameter priors $\alpha_i\in[0,3]$. The best-fit coefficients $\alpha_i$ are summarized in Tab.~\ref{tabBezier} and the corresponding contour plots, made by modyfing a free Python code \citep{2016JOSS....1...46B}, are portayed in Fig.~\ref{contourplotBezier} of \ref{cont}.
\begin{table*}[t]
    \centering
    \setlength{\tabcolsep}{1.em}
    \renewcommand{\arraystretch}{1.3}
    \resizebox{\textwidth}{!}{
    \begin{tabular}{lrll}
    \hline\hline
    {Correlation} & {$N_{cal}$} & {$Y_i-Y(z_i)$} & {$\sigma^2_{Y_i}$}\\
    \hline
    {$E_p$--$E_{iso}$} & {65} & {$\log E_{p,i}-a\left[\log E_{iso,i}-52\right]-b$} &
    {$\sigma^2_{\log E_{p,i}}+a^2\sigma^2_{\log E_{iso,i}}+\sigma^2$} \\
    {$L_p$--$E_p$} & {54} & {$\log L_{p,i}-a\log E_{p,i}-b-52$} &
    {$\sigma^2_{\log L_{p,i}}+a^2\sigma^2_{\log E_{p,i}}+\sigma^2$} \\
    {$L_0$--$E_p$--$T$} & {126} & {$\log L_{0,i}-a\log{E_{p,i}}+\log T_i - b$} &
    {$\sigma^2_{\log L_{0,i}}+a^2\sigma^2_{\log E_{p,i}}+\sigma^2_{\log T_i}+\sigma^2$} \\
    {$L_X$--$T_X$--$L_p$} & {20} & {$\log {L_{X,i}}-a\log {T_{X,i}}-b\log L_{p,i}-c$} & {$\sigma^2_{\log L_{X,i}}+a^2\sigma^2_{\log T_{X,i}}+b^2\sigma^2_{\log L_{p,i}}+\sigma^2$} \\
    \hline
    {Correlation} & {$N_{cos}$} & {$\mu_i$} & {$\sigma^2_{\mu,i}$} \\
     \hline
    {$E_p$--$E_{iso}$} & {118} &
    {$\mu_0+\frac{5}{2}\left[a^{-1}(\log E_{p,i}-b)-\log{\left(\frac{S_{b,i}}{1+z_i}\right)}+52\right]$} &
    {$\frac{25}{4}\left[a^{-2}\sigma^2_{\log E_{p,i}}+\sigma^2_{\log{S_{b,i}}}+\sigma^2\right]$} \\
    {$L_p$--$E_p$} & {101} &
    {$\mu_0+\frac{5}{2}\left[a\log{E_{p,i}}-\log{F_{p,i}}+b+52\right]$} &
    {$\frac{25}{4}\left[a^2\sigma^2_{\log E_{p,i}}+\sigma^2_{\log F_{p,i}}+\sigma^2\right]$} \\
    {$L_0$--$E_p$--$T$} & {182} &
    {$\mu_0+\frac{5}{2}\left[a\log{E_{p,i}}-\log{T_i}-\log{F_{0,i}}+b\right]$} &
    {$\frac{25}{4}\left[\sigma^2_{\log F_{0,i}}+a^2\sigma^2_{\log Ep,i}+\sigma^2_{\log{T_i}}+\sigma^2\right]$} \\
    {$L_X$--$T_X$--$L_p$} & {50} &
    {$\mu_0+\frac{5}{2}\left[\frac{1}{1-b}(a\log T_{X,i}+b\log{F_{p,i}}-\log{F_{X,i}}+c)+\frac{\log{(1+z_i)}}{(1-\beta)^{-1}}\right]$} &
    {$\frac{25}{4}\left[\frac{1}{(1-b)^2}(a^2\sigma^2_{\log T_{X,i}}+b^2\sigma^2_{\log F_{p,i}}+\sigma^2_{\log{F_{X,i}}})+\sigma^2\right]$} \\
    \hline
    \end{tabular}
    }
    \caption{Definitions of the quantities entering the calibration and cosmological log-likelihood functions for all the GRB correlations considered in this work.}
    \label{compcal}
\end{table*}

Through the above results, GRB data sets can be calibrated and employed in the total MCMC fit to search for the model best-fit parameters.
We consider the total log-likelihood
\begin{equation}
\label{loglike}
\ln\mathcal{L}=\ln\mathcal{L}_G+\ln\mathcal{L}_S+\ln\mathcal{L}_B\,,
\end{equation}
where $\ln\mathcal{L}_G$, $\ln\mathcal{L}_S$ and $\ln\mathcal{L}_B$ are the log-likelihood functions of GRB, Pantheon SNe Ia and BAO, respectively.

Below we describe each contribution to Eq.~\eqref{loglike}
\begin{enumerate}
\item \textbf{GRB log-likelihood}. It is composed of two parts
\begin{equation}
\ln\mathcal{L}_G=\ln\mathcal{L}_G^{cal}+\ln\mathcal{L}_G^{cos}\,.
\end{equation}
The calibration log-likelihood $\ln\mathcal{L}_G^{cal}$ determines the correlation coefficients by means of a \textit{calibrator sub-sample} of $N_{cal}$ GRB with $z\leq z_O$. For each correlation
Eqs.~\eqref{Eiso}, \eqref{Lp}, \eqref{L0} and \eqref{LpD}--\eqref{Lx} are calibrated with the luminosity distance $D_L^{cal}$ in Eq.~\eqref{callumdist}.
This log-likelihood reads
\begin{equation}\label{grbcal}
\ln\mathcal{L}_G^{cal}=-\frac{1}{2}\sum^{N_{cal}}_{i=1}\left\{\left[\frac{Y_i-Y(z_i)}{\sigma_{Y_i}}\right]^2+\ln(2\pi\sigma^2_{Y_i})\right\}.
\end{equation}
The definitions of $Y_i-Y(z_i)$ and $\sigma_{Y_i}$ in Eq.~\eqref{grbcal}, for each GRB correlation, are summarized in Tab.~\ref{compcal}.

On the other hand, the cosmological log-likelihood $\ln\mathcal{L}_G^{cos}$ determines the cosmological model parameters through the whole uncalibrated GRB data set
\begin{equation}\label{GRBcosm}
\ln\mathcal{L}_G^{cos}=-\frac{1}{2}\sum^{N_{cos}}_{i=1}\left\{\left[\frac{\mu_i-\mu(z_i)}{\sigma_{\mu_i}}\right]^2+\ln\left(2\pi\sigma^2_{\mu_i}\right)\right\},
\end{equation}
where the usual definition of the distance moduli is
\begin{equation}
\mu(z)=25+5\log\left[\frac{d_L(z)}{{\rm Mpc}}\right].
\end{equation}
and, for a flat universe, the luminosity distance is
\begin{equation}
d_L(z,z_t,j_{t})=\frac{c}{100 h_0}(1+z)\int^{z}_{0} \frac{dz'}{\xi(z^\prime,z_t,j_{t})},
\end{equation}
with $\xi(z,z_t,j_{t})$ given by the DHE or the DDPE methods, and the Hubble constant given by $H_0=100h_0$~km/s/Mpc.
For each GRB correlation, $\mu_i$ and $\sigma_{\mu_i}$ are listed in Tab.~\ref{compcal}, where we used $\mu_0=25-2.5\log(4\pi k^2)=-100.2$, with $k$ converting from Mpc to cm. The uncertainties take into due account the extrascatter $\sigma$ \citep{2005physics..11182D}.

\item \textbf{SNe Ia log-likelihood}. The SNe Ia log-likelihood is
\begin{equation}
    \ln\mathcal{L}_S=-\frac{1}{2}\sum^{N_S}_{i=1}\left[\Delta\Xi_i^T \textbf{C}^{-1} \Delta\Xi_i+\ln(2\pi |C|)\right]\,,
\end{equation}
where ${N_S}=6$ are the values of $\xi_i^{-1}$ obtained from SNe Ia within the flat-universe assumption \citep{2018ApJ...853..126R}, which are totally equivalent to the complete Pantheon catalog \citep{2018ApJ...859..101S}, \textbf{C} is the covariant matrix, $|C|$ its determinant, and we have defined  $\Delta\Xi_i=\xi_i^{-1}-\xi(z_i)^{-1}$.

\item \textbf{BAO log-likelihood}. The BAO log-likelihood is
\begin{equation}
    \ln\mathcal{L}_B=-\frac{1}{2}\sum^{N_B}_{i=1}\left\{\left[\frac{\Delta_i-\Delta(z_i)}{\sigma_{{\Delta_i}}}\right]^2+\ln(2\pi\sigma^2_{\Delta_i})\right\},
\end{equation}
where we took $N_B=8$ measurements $\Delta_i$ from \citet{2021MNRAS.503.4581L}, to be compared with the model values
\begin{equation}
    \Delta(z,z_t,j_{t})\equiv r_s\left[\frac{100h_0\xi(z,z_t,j_{t})}{cz}\right]^\frac{1}{3}\left[\frac{(1+z)}{d_L(z,z_t,j_{t})}\right]^\frac{2}{3}\,,
\end{equation}
where $r_s=(147.21\pm 0.48)$~Mpc is the comoving sound horizon at the drag epoch \citep{planck2018}. In our analysis we excluded the correlated WiggleZ data \citep{wigglez} to avoid the explicit dependence upon the matter density $\Omega_m$.

\end{enumerate}

\subsection{Numerical results for the DHE method}

The MCMC best-fit correlation (of each calibrated GRB data set) and cosmographic parameters from the DHE method, obtained by maximizing the log-likelihood in Eq.~\eqref{loglike}, are summarized in the upper part of Tab.~\ref{tabmodels} and portayed in the contour plots in Figs.~\ref{cont_A_DHE}--\ref{cont_D_DHE} of \ref{cont}.
\begin{table*}
\setlength{\tabcolsep}{0.8em}
\renewcommand{\arraystretch}{1.3}
\resizebox{\textwidth}{!}{
    \begin{tabular}{lcccccccc}
    \hline\hline
     {Correlation} & {$a$} & {$b$} & {$c$} & {$\sigma$} & {$h_0$} & {$z_{t}$}  & {$j_{t}$} \\
     \hline
     \multicolumn{8}{c}{DHE Model}\\
     \hline
     {$E_p-E_{iso}$} & {$0.750^{+0.056(0.096)}_{-0.067(0.095)}$} & {$1.769^{+0.093(0.138)}_{-0.083(0.141)}$} & {$-$} & {$0.298^{+0.039(0.065)}_{-0.032(0.050)}$} & {$0.689^{+0.016(0.025)}_{-0.013(0.021)}$} & {$0.716^{+0.138(0.268)}_{-0.104(0.164)}$} & {$1.04^{+0.25(0.41)}_{-0.27(0.41)}$}\\
     {$L_p-E_p$} & {$1.468^{+0.132(0.238)}_{-0.155(0.241)}$} & {$-3.320^{+0.329(0.549)}_{-0.369(0.637)}$} & {$-$} & {$0.349^{+0.069(0.123)}_{-0.064(0.094)}$} & {$0.693^{+0.013(0.022)}_{-0.016(0.024)}$} & {$0.684^{+0.136(0.253)}_{-0.092(0.146)}$} & {$1.09^{+0.28(0.42)}_{-0.25(0.41)}$}\\
     {$L_0-E_p-T$} & {$0.813^{+0.082(0.143)}_{-0.079(0.137)}$} & {$49.66^{+0.20(0.35)}_{-0.21(0.36)}$} & {$-$} & {$0.396^{+0.037(0.056)}_{-0.029(0.047)}$} & {$0.690^{+0.015(0.025)}_{-0.015(0.025)}$} & {$0.647^{+0.116(0.199)}_{-0.085(0.135)}$} & {$1.18^{+0.23(0.42)}_{-0.29(0.45)}$}\\
     {$L_X-T_X-L_p$} & {$-0.920^{+0.227(0.331)}_{-0.190(0.357)}$} & {$0.117^{+0.162(0.271)}_{-0.184(0.313)}$} & {$45.6^{+9.9(17.1)}_{-8.6(14.5)}$} & {$0.600^{+0.089(0.168)}_{-0.079(0.123)}$} & {$0.692^{+0.015(0.024)}_{-0.015(0.024)}$} & {$0.672^{+0.107(0.199)}_{-0.097(0.142)}$} & {$1.17^{+0.23(0.41)}_{-0.28(0.42)}$}\\
     \hline
     \multicolumn{8}{c}{DDPE Model}\\
     \hline
     {$E_p-E_{iso}$} & {$0.743^{+0.061(0.101)}_{-0.061(0.093)}$} & {$1.766^{+0.091(0.143)}_{-0.083(0.146)}$} & {$-$} &
     {$0.296^{+0.041(0.070)}_{-0.070(0.047)}$} & {$0.718^{+0.038(0.054)}_{-0.068(0.114)}$} & {$0.818^{+0.127(0.230)}_{-0.105(0.155)}$} & {$1.05^{+0.21(0.32)}_{-0.19(0.30)}$}\\
     {$L_p-E_p$} & {$1.465^{+0.138(0.235)}_{-0.142(0.243)}$} & {$-3.33^{+0.32(0.56)}_{-0.37(0.61)}$} & {$-$} & {$0.355^{+0.063(0.117)}_{-0.070(0.103)}$} & {$0.723^{+0.036(0.052)}_{-0.063(0.110)}$} & {$0.810^{+0.113(0.190)}_{-0.112(0.168)}$} & {$1.08^{+0.21(0.34)}_{-0.18(0.30)}$}\\
     {$L_0-E_p-T$} & {$0.807^{+0.086(0.141)}_{-0.074(0.134)}$} & {$49.69^{+0.19(0.34)}_{-0.22(0.37)}$} & {$-$} & {$0.401^{+0.032(0.054)}_{-0.031(0.050)}$} & {$0.708^{+0.044(0.062)}_{-0.069(0.116)}$} & {$0.761^{+0.113(0.217)}_{-0.096(0.142)}$} & {$1.06^{+0.19(0.32)}_{-0.22(0.33)}$}\\
     {$L_X-T_X-L_p$} & {$-0.893^{+0.200(0.325)}_{-0.246(0.391)}$} & {$0.126^{+0.162(0.231)}_{-0.179(0.323)}$} & {$45.2^{+9.6(17.1)}_{-8.7(12.7)}$} & {$0.599^{+0.103(0.156)}_{-0.076(0.113)}$} & {$0.694^{+0.017(0.025)}_{-0.015(0.024)}$} & {$0.765^{+0.119(0.225)}_{-0.104(0.150)}$} & {$1.11^{+0.22(0.34)}_{-0.22(0.37)}$}\\
     \hline
    \end{tabular}
    }
    \caption{Best-fit correlation parameters, of each calibrated GRB data set, and cosmographic parameters from DHE (upper part results) and DDPE (lower part results) methods. Each value is given with $2$ significant figures determined by its attached $1$--$\sigma$ errors; the attached $2$--$\sigma$ errors are also indicated in the round brackets.}
    \label{tabmodels}
\end{table*}

Focusing on the correlation parameters, all the values in Tab.~\ref{tabmodels} are consistent with the literature, \citep[see, for example,][]{2023MNRAS.523.4938M, 2020A&A...641A.174L}. The only exception is the $L_X$--$T_X$--$L_p$ correlation, for which the values in Tab.~\ref{tabmodels} differ from those reported in \citet{2022MNRAS.512..439C}, where, however, the constraints come from an uncalibrated correlation.

All the correlations provide results in agreement with the $\Lambda$CDM model \citep{planck2018}. In particular:
\begin{itemize}
    \item all the inferred $h_0$ are compatible within $1$-sigma confidence level (CL) with $h_0^P = 0.674\pm 0.005$ given by \citet{planck2018}, being the $E_p$--$E_{iso}$ and the $L_0$--$E_p$--$T$ correlations the ones providing values closer to $h_0^P$;
    \item these inferred $h_0$ are all incompatible at $>2$ sigma CL with the value inferred from SNe Ia measurements, $h_0^S=0.730\pm0.010$, given by \citet{2022ApJ...934L...7R};
    \item all the inferred transition redshifts are compatible within $1$-sigma CL with $z_t=0.632$, with the $L_0$--$E_p$--$T$ correlation providing the closest value;
    \item all the values of $j_t$ are compatible within $1$-sigma CL with $j=1$, being the $E_p$--$E_{iso}$ and the $L_p$--$E_p$ correlations the closest ones.
\end{itemize}

\subsection{Numerical results for the DDPE method}

The MCMC best-fit correlation (of each calibrated GRB data set) and cosmographic parameters from the DHE method, obtained by maximizing the log-likelihood in Eq.~\eqref{loglike}, are summarized in the lower part of Tab.~\ref{tabmodels} and portayed in the contour plots in Figs.~\ref{cont_A_DDPE}--\ref{cont_D_DDPE} of \ref{cont}.

Also in this case, only the results of the $L_X$--$T_X$--$L_p$ correlation (see Tab.~\ref{tabmodels}) differ from those in \citet{2022MNRAS.512..439C}, where the constraints were obtained from the uncalibrated correlation.

The correlation parameters obtained from this method are in agreement with those got from the DHE method, confirming that the $L_X-T_X-L_p$ correlation has the largest extrascatter.
However the cosmographic parameters differs from the previous case and we get discordant results with respect to the $\Lambda$CDM predictions \citep{planck2018}. In summary:
\begin{itemize}
    \item $E_p$--$E_{iso}$, $L_p$--$E_p$, and $L_0$--$E_p$--$T$ correlations provide large constraints on $h_0$ that are consistent at $1$-sigma CL with both $h_0^P$ and $h_0^S$;
    \item the $L_X-T_X-L_p$ correlation is the only one that is compatible within $1$-sigma CL with $h_0^P$ and incompatible at $>2$ sigma CL with $h_0^S$;
    \item the $L_X-T_X-L_p$ and the $L_0$--$E_p$--$T$ are the only correlations that give transition redshift values compatible within $2$-sigma CL with the concordance model;
    \item all the values of $j_t$ are compatible within $1$-sigma CL with $j=1$, being the $E_p$--$E_{iso}$ and the $L_0$--$E_p$--$T$ correlations the closest ones.
\end{itemize}

\section{Theoretical implications}\label{sec6}

We study the theoretical implications of the results in Tab.~\ref{tabmodels}.

First, we investigate the dark energy behavior to ascertain whether it acts as a cosmological costant or evolves with time.

Then, we determine \textit{a posteriori} the matter density $\Omega_m$ at each transition redshifts and check the compatibility with the value $\Omega_m^P=0.315\pm 0.007$ from \citet{planck2018}.

\subsection{Dark energy behavior}

To reconstruct the dark energy behavior, we follow the approach proposed in \citet{2023MNRAS.523.4938M}. We consider the Hubble rate with the hypothesis of spatially flat universe
\begin{equation}\label{hrate}
    H(z)=H_0\sqrt{\Omega_m(1+z)^3+\Omega_{de}F(z)},
\end{equation}
where $\Omega_{de}\equiv 1-\Omega_m$ and $F(z)\rightarrow1$ as $z\rightarrow0$. Further, $\Omega_{de}F(z) \gtrsim \Omega_m(1+z)^3$ for $z\rightarrow 0$. These conditions ensure that dark energy dominates over matter at late-times and that $H=H_0$ at $z=0$.

\begin{table*}[t]
    \centering
    \setlength{\tabcolsep}{2em}
    \renewcommand{\arraystretch}{1.3}
    \resizebox{\textwidth}{!}{
    \begin{tabular}{llccccc}
    \hline {Model} & {Correlation} & {$F_t$} & {$F^\prime_t$} & {$F^{\prime\prime}_t$} & {$\Omega_m$} \\ \hline
     {DHE} & {$E_p-E_{iso}$} & {$0.957^{+0.210(0.310)}_{-0.178(0.261)}$} & {$-0.412^{+1.061(1.788)}_{-0.858(1.267)}$} & {$-0.191^{+1.236(2.046)}_{-1.153(1.726)}$} & {$0.341^{+0.242(0.412)}_{-0.194(0.290)}$} \\
      {DDPE} & {} & {$0.652^{+0.478(0.679)}_{-0.754(1.216)}$} & {$-1.463^{+1.630(2.501)}_{-2.081(3.232)}$} & {$-0.782^{+1.306(1.964)}_{-1.567(2.455)}$} & {$0.397^{+0.276(0.427)}_{-0.352(0.549)}$}\\ \hline
      {DHE} & {$L_p-E_p$}& {$0.924^{+0.177(0.275)}_{-0.198(0.278)}$} & {$-0.351^{+0.962(1.615)}_{-0.831(1.202)}$} & {$-0.049^{+1.260(1.999)}_{-1.116(1.719)}$} & {$0.338^{+0.231(0.393)}_{-0.198(0.290)}$}\\
      {DDPE} & {} & {$0.609^{+0.448(0.636)}_{-0.697(1.160)}$} & {$-1.512^{+1.500(2.224)}_{-1.993(3.173)}$} & {$-0.760^{+1.241(1.876)}_{-1.494(2.423)}$} & {$0.400^{+0.254(0.380)}_{-0.338(0.541)}$}\\ \hline
      {DHE} & {$L_0-E_p-T$} & {$0.955^{+0.180(0.275)}_{-0.177(0.269)}$} & {$-0.147^{+0.864(1.366)}_{-0.742(1.113)}$} & {$0.291^{+1.146(1.939)}_{-1.193(1.828)}$} & {$0.325^{+0.226(0.362)}_{-0.193(0.294)}$}\\
      {DDPE} & {} & {$0.769^{+0.495(0.697)}_{-0.722(1.173)}$} & {$-0.977^{+1.560(2.409)}_{-1.940(3.039)}$} & {$-0.503^{+1.341(2.095)}_{-1.671(2.596)}$} & {$0.374^{+0.305(0.442)}_{-0.381(0.600)}$}\\ \hline
      {DHE} & {$L_X-T_X-L_p$} & {$0.935^{+0.187(0.279)}_{-0.186(0.272)}$} & {$-0.283^{+0.860(1.400)}_{-0.819(1.165)}$} & {$0.173^{+1.109(1.880)}_{-1.193(1.748)}$} & {$0.334^{+0.212(0.349)}_{-0.201(0.289)}$}\\
      {DDPE} & {} & {$0.902^{+0.237(0.335)}_{-0.215(0.312)}$} & {$-0.728^{+1.073(1.723)}_{-0.964(1.355)}$} & {$-0.225^{+1.118(1.748)}_{-1.055(1.618)}$} & {$0.360^{+0.217(0.352)}_{-0.194(0.275)}$}\\ \hline
    \end{tabular}
    }
    \caption{Constraints on the dark energy behaviour via $F_t$, $F^\prime_t$ and $F^{\prime\prime}_t$, and on the matter density $\Omega_m$, for each GRB correlation, using DHE and DDPE methods.}
    \label{tabF1F2Om}
\end{table*}

To find the deceleration and the jerk parameters, we use the first of Eqs.~\eqref{Hderiv} and Eq.~\eqref{dq} together with Eq.~\eqref{hrate}, and obtain
\begin{align}
\label{q}
q(z)=&\,-1+\frac{(1+z)\left[3\Omega_m(1+z)^2+\Omega_{de}F^\prime(z)\right]}{2\left[\Omega_m(1+z)^3+\Omega_{de}F(z)\right]}\,,\\
\label{j}
j(z)=&\,1+\frac{\Omega_{de}(1+z)\left[-2F^\prime(z)+(1+z)F^{\prime\prime}(z)\right]}{2\left[\Omega_m(1+z)^3+\Omega_{de}F(z)\right]}\,.
\end{align}
Imposing $z=z_t$, we explicit $F^\prime_t$ and $F^{\prime\prime}_t$ from Eq.~\eqref{q}--\eqref{j}
\begin{subequations}\label{F1}
    \begin{align}
        F^\prime_t&=-\frac{\Omega_m(1+z_t)^2}{1-\Omega_m}+\frac{2F_t}{1+z_t},&\\
         F^{\prime\prime}_t&=-\frac{2\Omega_m(2-j_t)(1+z_t)}{1-\Omega_m}+\frac{2(1+j_t)F_t}{(1+z_t)^2}.&
    \end{align}
\end{subequations}
Eqs.~\eqref{F1} are insufficient to constraint the evolution of dark energy, since an expression for $F_t$ is missing. To derive it we substitute inside Eq.~\eqref{hrate} the B\'ezier interpolation at $z_t$ and require that $H_{2,t}\equiv H_2(z_t)\equiv H_t$. In this way we get
\begin{subequations}\label{Ft}
    \begin{align}
        F_t&=\frac{H_{2,t}^2-H_0^2\Omega_m(1+z_t)^3}{H_0^2(1-\Omega_m)},& \\
        F^\prime_t&=\frac{2H_{2,t}^2-3H_0^2\Omega_m(1+z_t)^3}{H_0^2(1-\Omega_m)},&\\
        F^{\prime\prime}_t&=\frac{2H_{2,t}^2(1+j_t)-6H_0^2\Omega_m(1+z_t)^3}{H_0^2(1-\Omega_m)(1+z_t)^2}.&
    \end{align}
\end{subequations}

Substituting in Eqs.~\eqref{Ft} the values from Tab.~\ref{tabmodels} and $\Omega_m$ from \citet{planck2018}, we obtain the constraints on $F_t$ and derivatives listed in Tab.~\ref{tabF1F2Om}.

Within both DHE and DDPE methods, all the correlations provide values of $F_t$, $F^\prime_t$ and $F_t^{\prime\prime}$ that are compatible with the $\Lambda$CDM model, for which $F_t=1$ and  $F^\prime_t=F_t^{\prime\prime}=0$. In particular, the results of the $E_p-E_{iso}$ correlation are in agreement with the ones found in \citet{2023MNRAS.523.4938M}.

Looking more in details at the best fit values of Tab.~\ref{tabF1F2Om}, the GRB correlations providing values of $F_t$ and derivatives that better reproduce the behavior of the cosmological constant are:
\begin{itemize}
    \item the $L_0-E_p-T$ correlation for the DHE model, and
    \item the $L_X-T_X-L_p$ correlation for the DDPE model.
\end{itemize}

In conclusion, if the dark energy evolves with time, this evolution is very slow and almost undistinguishable from the case of the cosmological constant.

\subsection{Matter density behavior}

To constrain the matter density $\Omega_m$, we substitute Eq.~\eqref{hrate} and its derivative into the first of Eqs.~\eqref{Hderiv} and obtain
\begin{equation}\label{Om}
    \Omega_m=\frac{2F_t-F^\prime_t(1+z_t)}{2F_t-(1+z_t)[F^\prime_t-(1+z_t)^2]}\,
\end{equation}
that reduces to the concordance paradigm expectation when $F_t=1$ and $F^\prime_t=0$. Now, we substitute into Eq.~\eqref{Om} the values of $F_t$ and $F^\prime_t$, listed in Tab.~\ref{tabF1F2Om}, and the ones of $z_t$, listed in Tab.~\ref{tabmodels}. Finally, we get the matter density constraints, for DHE and DDPE methods and for each considered GRB data set, listed in the last column of Tab.~\ref{tabF1F2Om}.

We distinguish between DHE and DDPE methods.
\begin{itemize}
    \item Within the DHE method, all the values of $\Omega_m$ obtained from all GRB correlations are consistent within $1$-sigma CL with the value from \citet{planck2018}, in particular the closest values are obtained from the $L_0-E_p-T$ and the $L_X-T_X-L_p$ correlations.
    \item {Within the DDPE method, the values of $\Omega_m$ obtained from all GRB correlations are also consistent, as in the DHE case, with \citet{planck2018}. Moreover, the higher values of $\Omega_m$ found for the DDPE method are compatible with results obtained by analyzing the cosmic evolution using GRB, see for example \citep{2021MNRAS.503.4581L,2021JCAP...09..042K, 2021ApJ...908..181M}.}
\end{itemize}

\section{Final outlooks and perspectives}\label{sec7}

In this work we extend what was done in \citet{2023MNRAS.523.4938M} by using, beside the $E_p-E_{iso}$ correlations, other three GRB correlation functions, namely the $L_0-E_p-T$, $L_p-E_p$ and $L_X-T_X-L_p$ correlations to investigate the transition epoch at which dark energy begins to dominate over matter and the universe begins to accelerate. To do so, we consider a model-independent approach employing two methods already used in the literature with promising results, the DHE and DDPE methods \citep{capozziellodunsbyluongo}. The first consists in the direct expansion of the Hubble parameter around the redshift at which the transition begins, $z_t$, while the latter focuses in expanding the deceleration parameter around $z_t$ and then deriving the Hubble rate. Considering that the GRB correlations suffer from the \textit{circularity} problem \citep{luongomuccino2021}, making them model-dependent, we interpolate the OHD sample with B\'ezier polynomials to build the correlations in a model-independent way. Afterwards, we use a MCMC simulation employing the Metropolis-Hastings algorithm written in Python to get constraints on the coefficients inside the correlations and on the cosmographic parameters, \textit{i.e.} the reduced Hubble constant $h_0$, the transition redshift $z_t$ and the jerk parameter at the transition, $j_t$ for both the DHE and DDPE methods. The results we obtained are shown in Tab.~\ref{tabmodels}, whereas the corresponding contours are shown in Figs.~\ref{cont_A_DHE}--\ref{cont_D_DDPE}.

What we inferred using the DHE method is that the correlation parameters of each GRB dataset are consistent with the ones found in the literature with the $L_X$--$T_X$--$L_p$ correlation being the one with the highest extrascatter term. For what concerns the cosmographic parameters, the reduced Hubble constant $h_0$ from our computations is compatible within $1$--$\sigma$ with the value from the \citet{planck2018}. In particular, the $E_p$--$E_{iso}$ and $L_0$--$E_p$--$T$ correlations are the ones with the closest value to $h_0^P$. Moreover, the results we obtained for $h_0$ are all incompatible at $>2$ sigma CL with the value found using SNe Ia in \cite{2022ApJ...934L...7R}.

Regarding the redshift at the transition, $z_t$, all the values inferred from our simulations are compatible within $1$--$\sigma$ with the value expected from the concordance paradigm. The $L_0$--$E_p-T$ correlation is the one that provides a closer value with the $z_t$ from the $\Lambda$CDM framework.

Finally, all the values of the jerk parameter at the transition, $j_t$, are compatible within $1$--$\sigma$ with the expectation from the $\Lambda$CDM scenario, \textit{i.e.} $j=1$ with in particular two correlations, $E_p$--$E_{iso}$ and $L_p$--$E_p$, being the ones closer to $j=1$.

Using the DDPE method we found that the correlation parameters are in agreement the ones found using the DHE method confirming again the highest value of the extrascatter term for the $L_X$--$T_X$--$L_p$ correlation. Moving to the cosmographic parameters the results we got with this method differ from the ones we got using the DHE methodology and are discordant with the predictions of the concordance paradigm. In particular, for what concerns $h_0$ we found that the correlations that are consistent with $1$--$\sigma$ with both the reduced Hubble constant from \cite{planck2018} and \cite{2022ApJ...934L...7R} are the $E_p$--$E_{iso}$, $L_p$--$E_p$ and $L_0$--$E_p$--$T$ while the $L_X$--$T_X$--$L_p$ correlation is the only one which is not compatible at $>2$ sigma CL with $h_0^S$ but compatible at $1$--$\sigma$ with $h_0^P$.

For what concerns the transition redshift we have a compatibility with the cosmological model at $2$--$\sigma$ only for two out of four correlations, namely the $L_X$--$T_X$--$L_p$ and $L_0$--$E_p$--$T$. Also, the values we inferred for the jerk parameter using the DDPE method are compatible for all the correlation functions with $j=1$ at $1$--$\sigma$, with the closest compatibility for the $E_p$--$E_{iso}$ and $L_0$--$E_p$--$T$ correlations.

Afterwards, we investigate the behaviour and evolution of a generic form of dark energy to check if it is compatible with the predictions of the $\Lambda$CDM of a cosmological constant or it can evolve. Considering the DHE and DDPE methodology for all of the four GRB correlation functions we used in this work we assert that for $F_t$ all four correlations have a compatibility with the attended result from the cosmological paradigm, \textit{i.e.} $F_t=1$, for both the DHE and DDPE methods at both $1$--$\sigma$ and $2$--$\sigma$. Concerning the first and second derivative of $F_t$ we have an overall compatibility with the values expected for the concordance model, \textit{i.e.} $F^\prime_t=F^{\prime\prime}_t=0$. In particular, the correlations that better reproduce the behaviour of dark energy as a cosmological constant are the $L_0$--$E_p$--$T$ correlation for the DHE method and the $L_X$--$T_X$--$L_p$ for the DDPE method. In conclusion, if an evolution is plausible dark energy should evolve in a slowly way in order to not be distinguishable from the case of the cosmological constant $\Lambda$. Finally, we also compared our results for the $E_p-E_{iso}$ with the ones got in \citet{2023MNRAS.523.4938M} finding an overall compatibility.

We also check \textit{a posteriori} if the matter density $\Omega_m$ inferred for the case of a generic dark energy term is compatible or not with the value found by the \citet{planck2018}. In order to do that, we substituted the values we found for the transition redshift for both the DHE and DDPE methods and the ones found for $F_t$ and $F^\prime_t$. Regarding the values inferred from the DHE method, all the values we obtained for the matter density for all the GRB correlation functions are in agreement at $1$--$\sigma$ with the one from \citet{planck2018}. Specifically, the values which are the closest with $\Omega_m^P$ are the ones inferred from the $L_0$--$E_p$--$T$ and $L_X$--$T_X$--$L_p$ correlations. {For the DDPE method, we also find a compatibility with the value $\Omega_m^P$ at $2$-$\sigma$ even though for this method we find higher values of the matter energy density compatible, otherwise, with other results in the literature \cite{2021MNRAS.503.4581L,2021JCAP...09..042K, 2021ApJ...908..181M}.}

Future works will be focused on extending the use of B\'ezier polynomials. For example, using them to fit BAO measurements through their volume-averaged distance in order to calibrate GRB correlation functions in both cases of a universe with zero and non-zero curvature as done in \citet{2023MNRAS.518.2247L}.

\section*{Acknowledgements}
ACA and SC  acknowledge the Istituto Nazionale di Fisica Nucleare (INFN) Sez. di Napoli, Iniziativa Specifica QGSKY.
The work  by OL is  partially financed by the Ministry of Education and Science of the Republic of Kazakhstan, Grant: IRN AP19680128. The work  by MM is  partially financed by the Ministry of Education and Science of the Republic of Kazakhstan, Grant: IRN BR21881941. ACA is grateful to Mr. Sebastiano Tomasi for discussions on the numerical analysis.
This paper is based upon work from COST Action CA21136 {\it Addressing observational tensions in cosmology with systematics and fundamental physics} (CosmoVerse) supported by COST (European Cooperation in Science and Technology).

\newpage

\appendix
\onecolumngrid

\section{Contour plots}\label{cont}

\begin{figure}[H]
    \centering
    \includegraphics[width=0.80\hsize,clip]{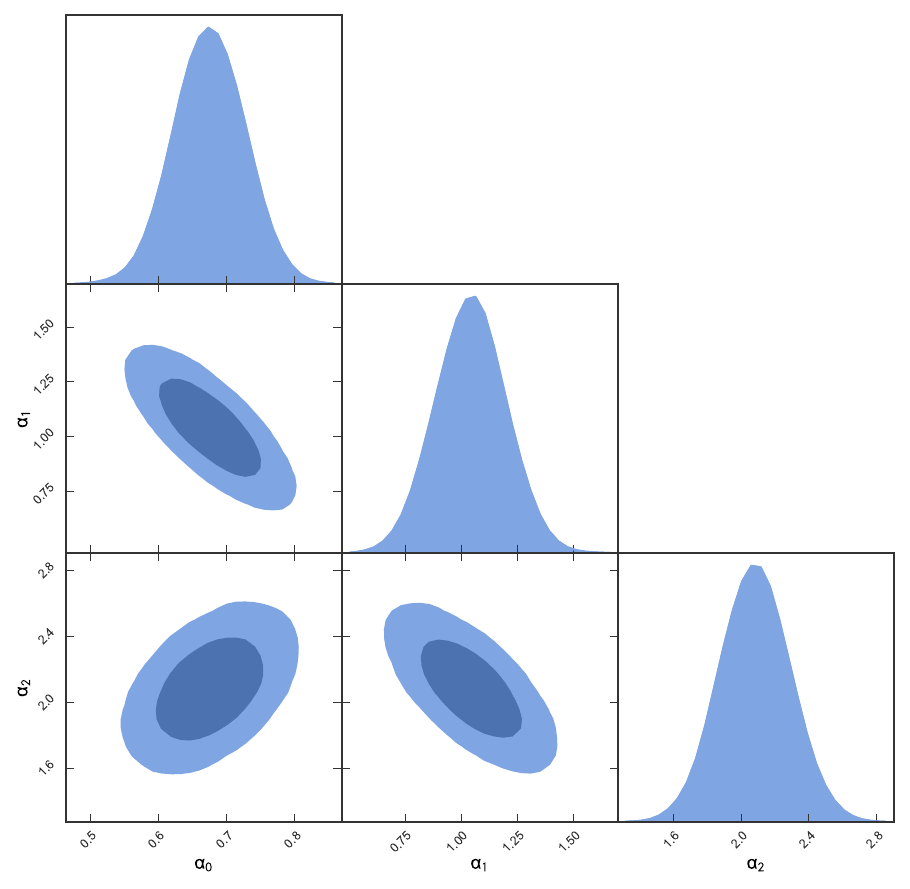}
    \caption{Contour plot of the best-fit coefficients $\alpha_i$ of the B\'ezier interpolation of the OHD catalog.}
    \label{contourplotBezier}
\end{figure}

Fig.~\ref{contourplotBezier} displays the contour plots of the MCMC best-fit coefficients $\alpha_i$ obtained from the B\'ezier interpolation of the OHD catalog, through the maximization of the log-likelihood in Eq.~\eqref{loglikeOHD}. The corresponding best-fit values and attached errors are summarized in Tab.~\ref{tabBezier}.

\begin{figure}[H]
    \centering
    \includegraphics[width=\hsize,clip]{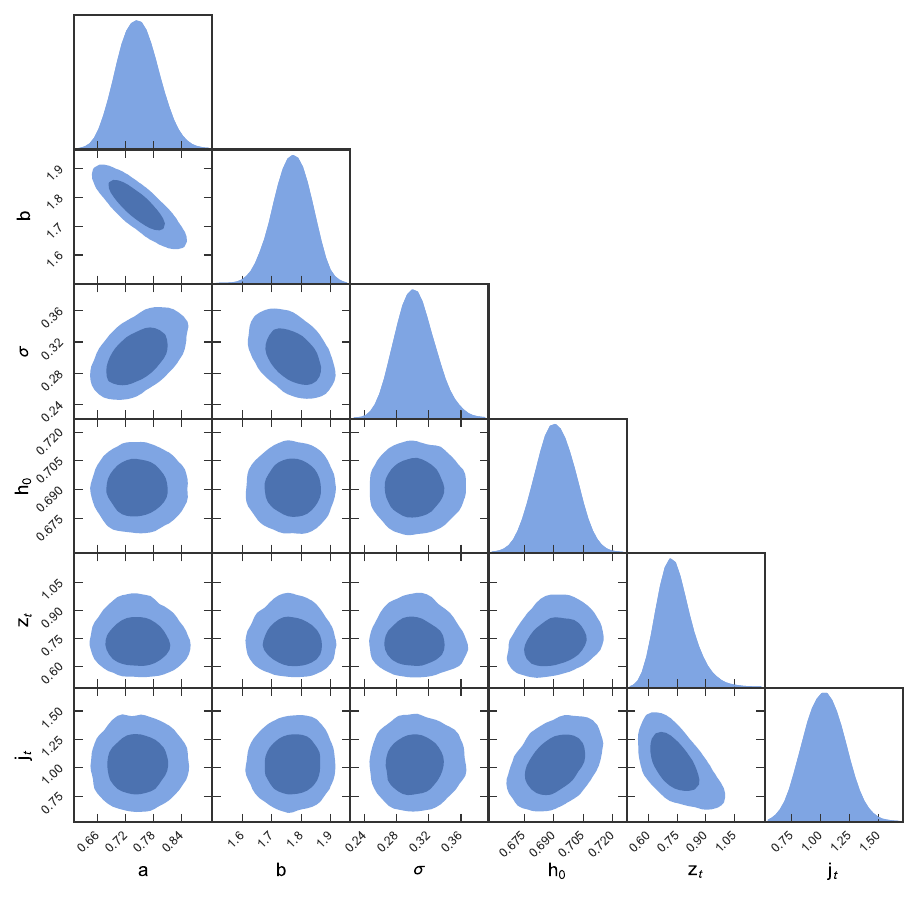}
    \caption{Contour plot of the best-fit parameters for the $E_p$--$E_{iso}$ correlation and the DHE method.}
    \label{cont_A_DHE}
\end{figure}

\begin{figure}[H]
    \centering
    \includegraphics[width=\hsize,clip]{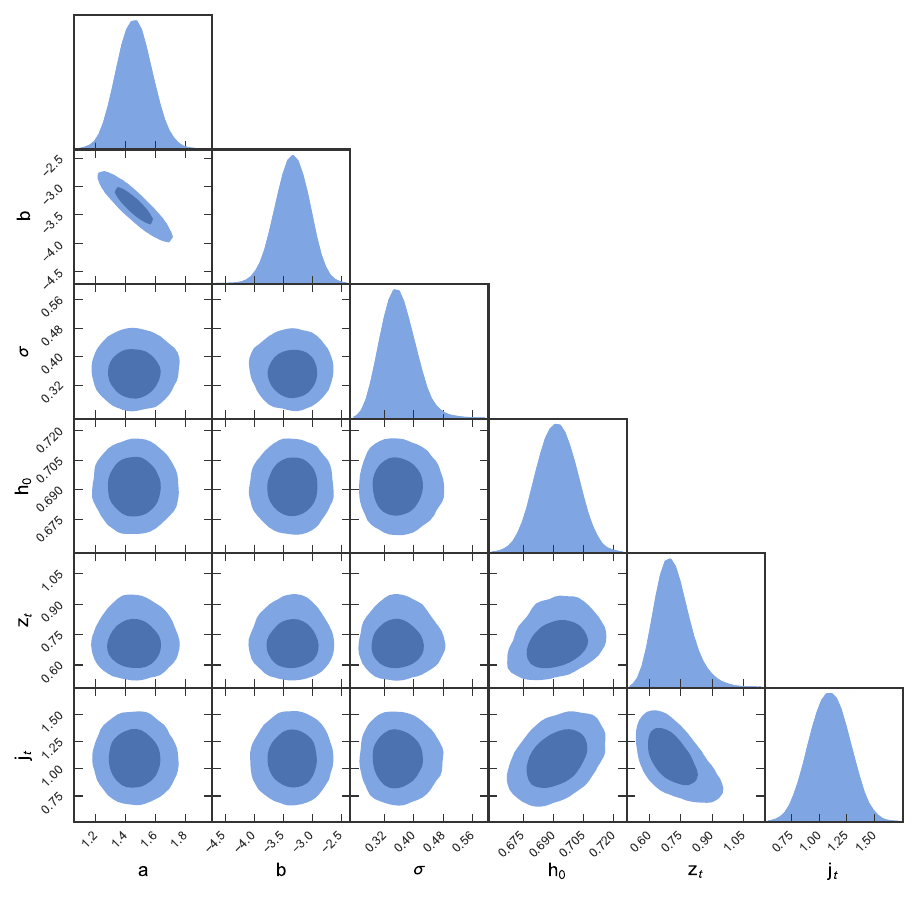}
    \caption{Contour plot of the best-fit parameters for the $L_p$--$E_p$ correlation and the DHE method.}
    \label{cont_Y_DHE}
\end{figure}

\begin{figure}[H]
    \centering
    \includegraphics[width=\hsize,clip]{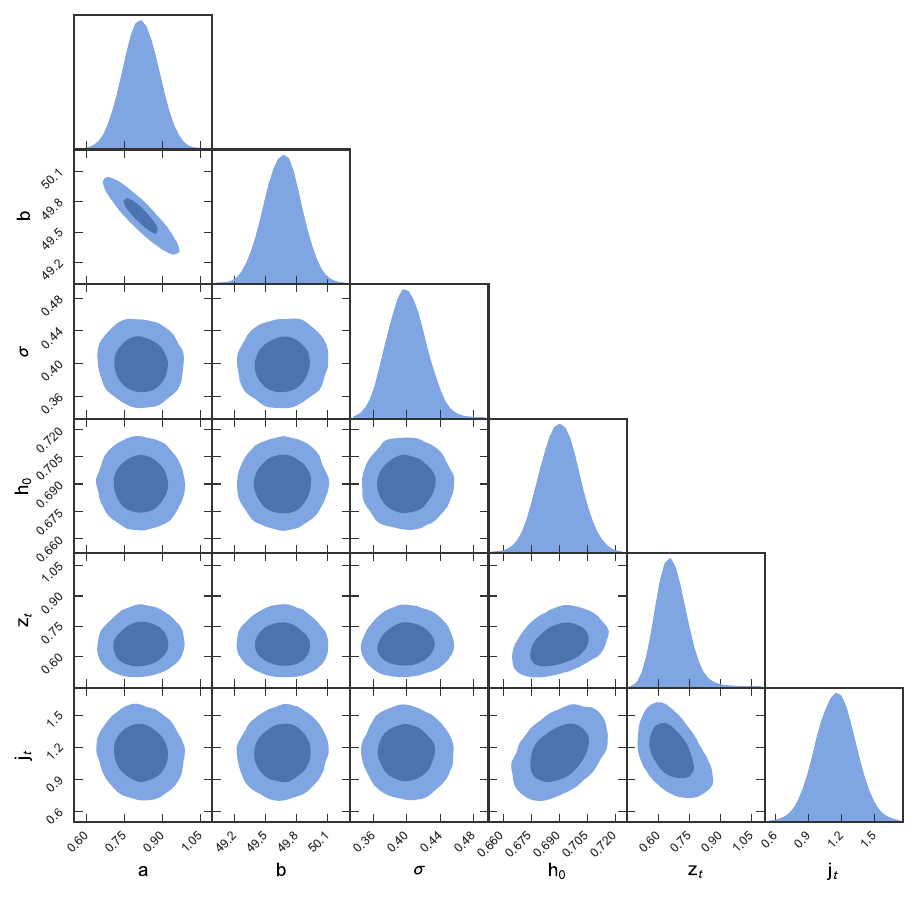}
    \caption{Contour plot of the best-fit parameters for the $L_0$--$E_p$--$T$ correlation and the DHE method.}
    \label{cont_C_DHE}
\end{figure}

\begin{figure}[H]
    \centering
    \includegraphics[width=\hsize,clip]{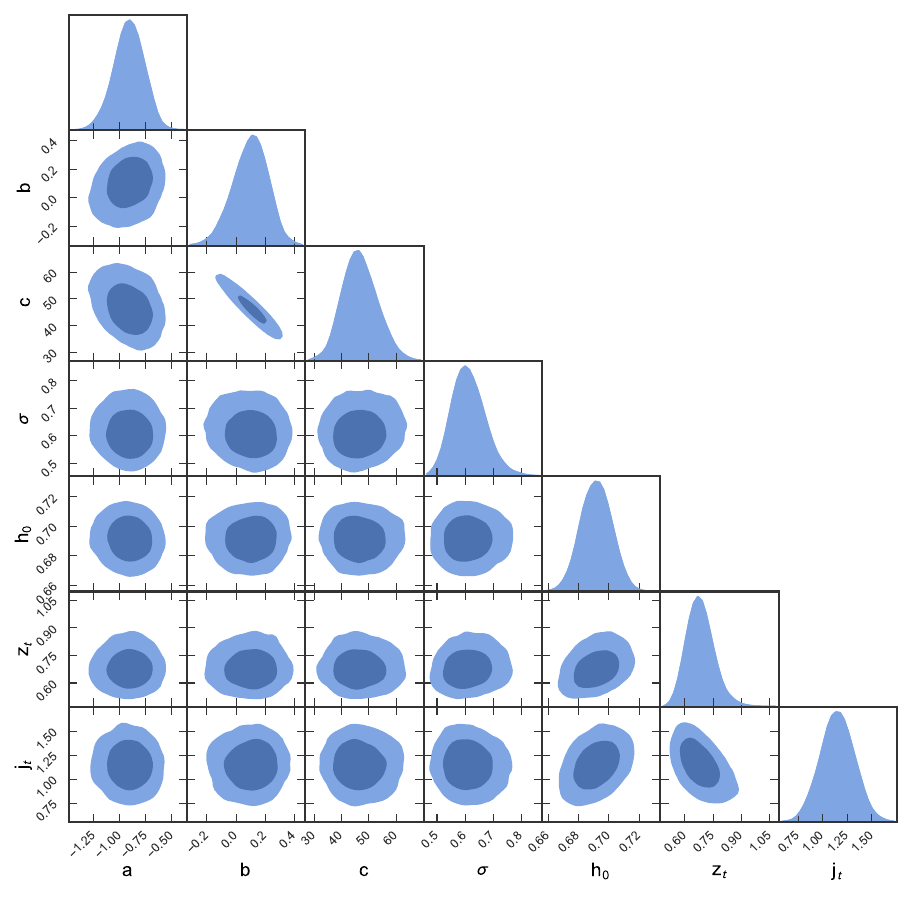}
    \caption{Contour plot of the best-fit parameters for the $L_X$--$T_X$--$L_p$ correlation and the DHE method.}
    \label{cont_D_DHE}
\end{figure}

Figs.~\ref{cont_A_DHE}, \ref{cont_Y_DHE}, \ref{cont_C_DHE} and \ref{cont_D_DHE} show
the contour plots of the MCMC best-fit correlation (of each calibrated GRB data set) and cosmographic parameters obtained by applying the DHE method in the maximization of the log-likelihood in Eq.~\eqref{loglike}. The correlation parameters displayed in these contour plots belong to the $E_p$--$E_{iso}$, the $L_p$--$E_p$, the $L_0$--$E_p$--$T$, and the $L_X$--$T_X$--$L_p$ correlations, respectively.
The corresponding best-fit values and attached errors are summarized in the upper part of Tab.~\ref{tabmodels}.

\begin{figure}[H]
    \centering
    \includegraphics[width=\hsize,clip]{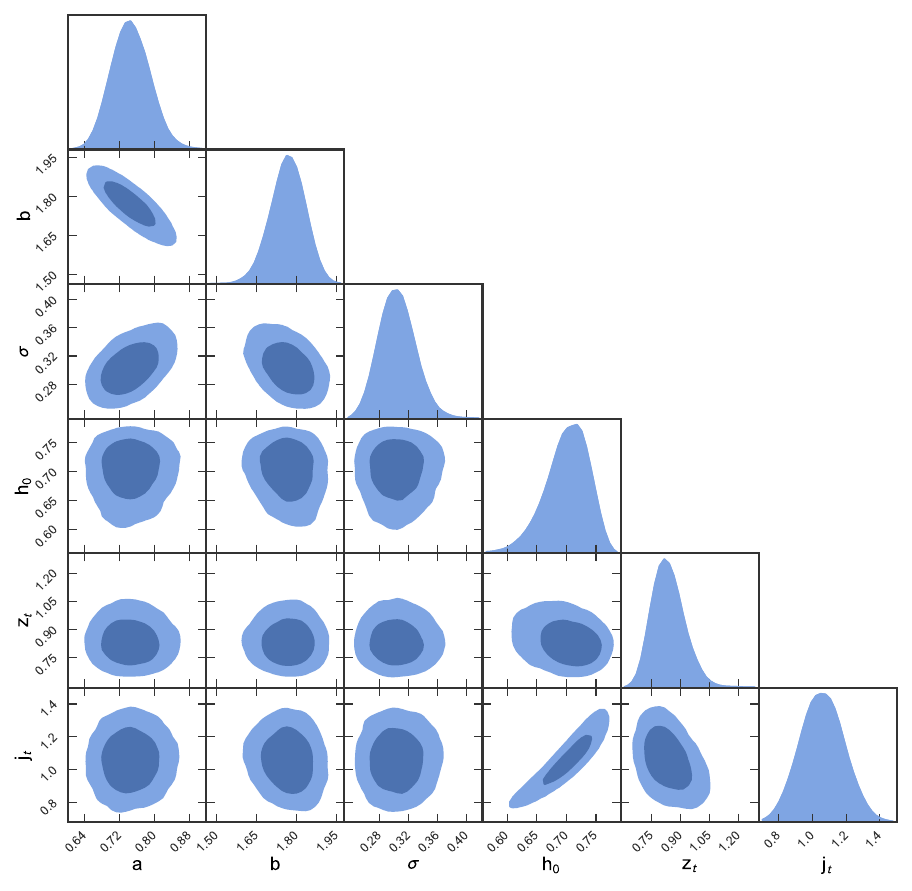}
    \caption{Contour plot of the best-fit parameters for the $E_p$--$E_{iso}$ correlation and the DDPE method.}
    \label{cont_A_DDPE}
\end{figure}

\begin{figure}[H]
    \centering
    \includegraphics[width=\hsize,clip]{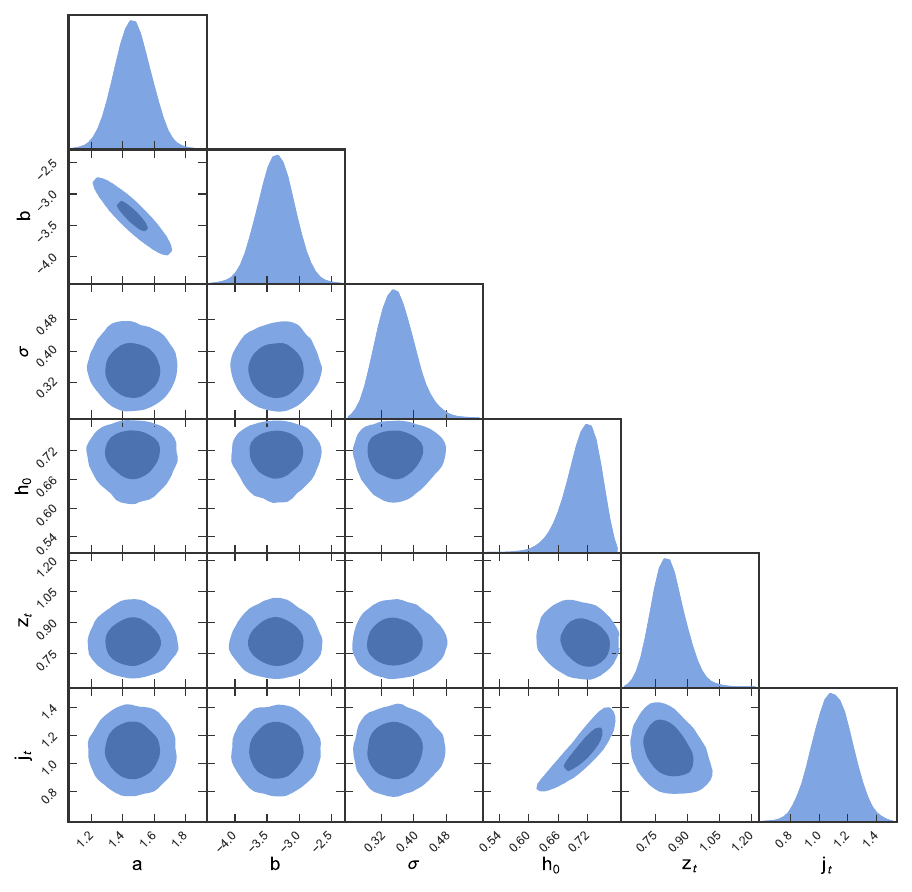}
    \caption{Contour plot of the best-fit parameters for the $L_p$--$E_p$ correlation and the DDPE method.}
    \label{cont_Y_DDPE}
\end{figure}

\begin{figure}[H]
    \centering
    \includegraphics[width=\hsize,clip]{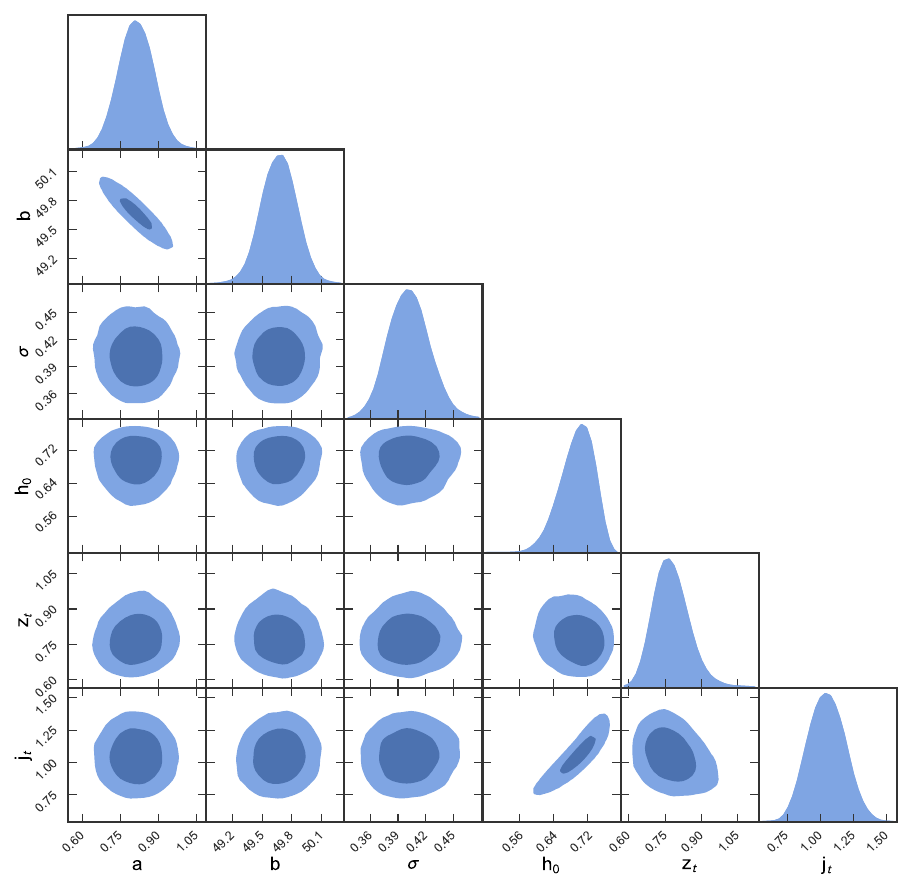}
    \caption{Contour plot of the best-fit parameters for the $L_0$--$E_p$--$T$ correlation and the DDPE method.}
    \label{cont_C_DDPE}
\end{figure}

\begin{figure}[H]
    \centering
    \includegraphics[width=\hsize,clip]{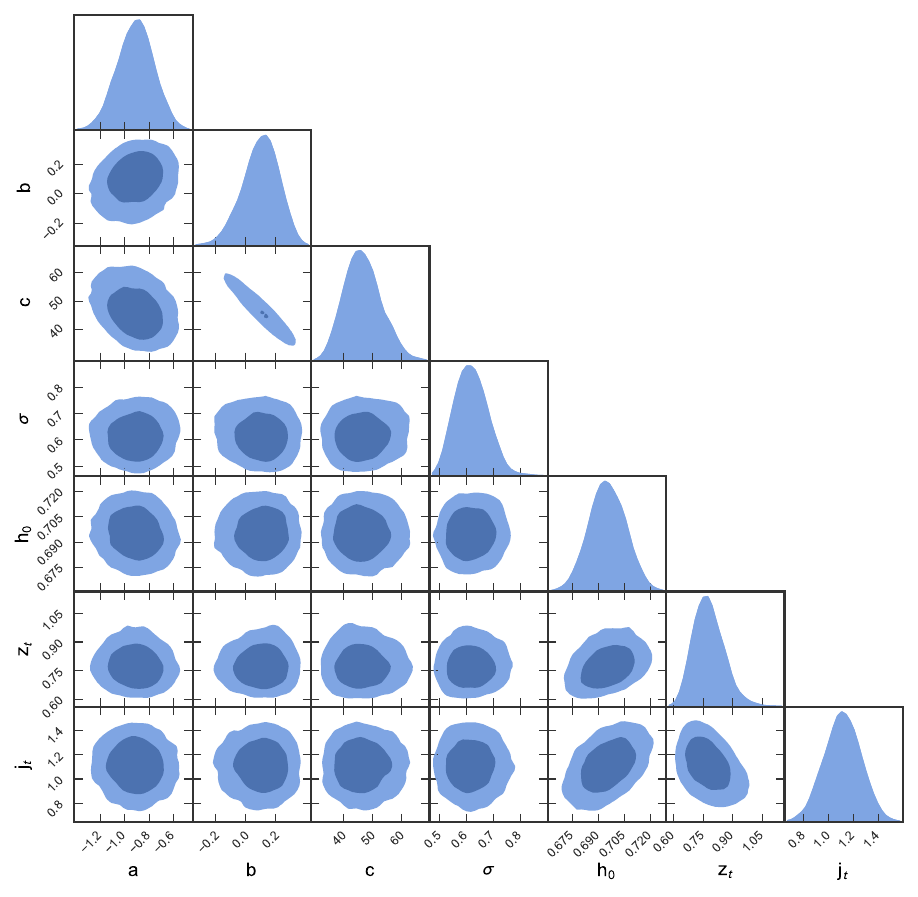}
    \caption{Contour plot of the best-fit parameters for the $L_X$--$T_X$--$L_p$ correlation and the DDPE method.}
    \label{cont_D_DDPE}
\end{figure}

Figs.~\ref{cont_A_DDPE}, \ref{cont_Y_DDPE}, \ref{cont_C_DDPE} and \ref{cont_D_DDPE} show the contour plots of the MCMC best-fit correlation (of each calibrated GRB data set) and cosmographic parameters obtained by applying the DDPE method in the maximization of the log-likelihood in Eq.~\eqref{loglike}. The correlation parameters displayed in these contour plots belong to the $E_p$--$E_{iso}$, the $L_p$--$E_p$, the $L_0$--$E_p$--$T$, and the $L_X$--$T_X$--$L_p$ correlations, respectively.
The corresponding best-fit values and attached errors are summarized in the lower part of Tab.~\ref{tabmodels}.

\section{Error propagation for $F_t$, $F^\prime_t$, $F^{\prime\prime}_t$ and $\Omega_m$}

To compute the errors of Tab.~\ref{tabF1F2Om}, we utilized the standard error propagation formula
\begin{equation}
\delta\sigma_i=\sum_j\Bigg\lvert\frac{\partial\sigma_i}{\partial y_j}\Bigg\lvert\delta y_i\,.
\end{equation}
where for $\sigma=\{F_t, F^\prime_t, F^{\prime\prime}_t\}$ we have $y=\{\Omega_m, H_0, z_t, j_t\}$, whereas for $\sigma=\Omega_m$ we have $y=z_t$.

For $F_t$, $F^\prime_t$ and $F^{\prime\prime}_t$ the error propagation formula leads to
\begin{subequations}
    \begin{align}
        \delta F_t=\,&\Bigg\lvert\frac{\partial F_t}{\partial\Omega_m}\Bigg\lvert\delta\Omega_m+\Bigg\lvert\frac{\partial F_t}{\partial H_0}\Bigg\lvert\delta H_0+\Bigg\lvert\frac{\partial F_t}{\partial z_t}\Bigg\lvert\delta z_t\,\\
        \delta F^\prime_t=\,&\Bigg\lvert\frac{\partial F^\prime_t}{\partial\Omega_m}\Bigg\lvert\delta\Omega_m+\Bigg\lvert\frac{\partial F^\prime_t}{\partial H_0}\Bigg\lvert\delta H_0+\Bigg\lvert\frac{\partial F^\prime_t}{\partial z_t}\Bigg\lvert\delta z_t\,,\\
        \delta F^{\prime\prime}_t=\,&\Bigg\lvert\frac{\partial F^{\prime\prime}_t}{\partial\Omega_m}\Bigg\lvert\delta\Omega_m+\Bigg\lvert\frac{\partial F^{\prime\prime}_t}{\partial H_0}\Bigg\lvert\delta H_0+\Bigg\lvert\frac{\partial F^{\prime\prime}_t}{\partial z_t}\Bigg\lvert\delta z_t+\Bigg\lvert\frac{\partial F^{\prime\prime}_t}{\partial j_t}\Bigg\lvert\delta j_t\,.
    \end{align}
\end{subequations}
We start writiting explicitly the partial derivatives of $F_t$
\begin{subequations}\label{36}
    \begin{align}
        &\frac{\partial F_t}{\partial\Omega_m}=\frac{H^2_{2,t}-H_0^2\Omega_m(1+z_t)^3-H_0^2(1-\Omega_m)(1+z_t)^3}{H_0^2(1-\Omega_m)^2}\,,\\
        &\frac{\partial F_t}{\partial H_0}=-\frac{2\left[H^2_{2,t}-H_0^2\Omega_m(1+z_t)^3\right]+2H_0^2\Omega_m(1+z_t)^3}{H_0^3(1-\Omega_m)}\,,\\
        &\frac{\partial F_t}{\partial z_t}=\frac{2H_{2,t}\frac{dH_{2,t}}{dz_t}-3H_0^2\Omega_m(1+z_t)^2}{H_0^2(1-\Omega_m)}\,.
    \end{align}
\end{subequations}
Then, we proceed with the partial derivatives of $F^\prime(z_t)$
\begin{subequations}\label{37}
    \begin{align}
        &\frac{\partial F^\prime_t}{\partial\Omega_m}=\frac{2H^2_2(z_t)-3H_0^2\Omega_m(1+z_t)^3-3H_0^2(1-\Omega_m)(1+z_t)^3}{H_0^2(1-\Omega_m)^2}\,,\\
        &\frac{\partial F^\prime_t}{\partial H_0}=-\frac{2\left[2H_2^2(z_t)-3H_0^2\Omega_m(1+z_t)^3\right]+6H_0^2\Omega_m(1+z_t)^3}{H_0^3(1-\Omega_m)}\,,\\
        &\frac{\partial F^\prime_t}{\partial z_t}=\frac{4H_2(z_t)\frac{dH_2(z_t)}{dz_t}-9H_0^2\Omega_m(1+z_t)^2}{H_0^2(1-\Omega_m)}\,.
    \end{align}
\end{subequations}
Finally, we deal with the partial derivatives of $F^{\prime\prime}_t$
\begin{subequations}\label{38}
    \begin{align}
        &\frac{\partial F^{\prime\prime}_t}{\partial\Omega_m}=\frac{2H^2_2(z_t)(1+j_t)-6H_0^2\Omega_m(1+z_t)^3-6H_0^2(1+z_t)^3(1-\Omega_m)}{H_0^2(1-\Omega_m)^2(1+z_t)^2}\,,\\
        &\frac{\partial F^{\prime\prime}_t}{\partial H_0}=-\frac{2\left[2H^2_2(z_t)(1+j_t)-6H_0^2\Omega_m(1+z_t)^3\right]+12H_0^2\Omega_m(1+z_t)^3}{H_0^3(1-\Omega_m)(1+z_t)^2}\,,\\
        &\frac{\partial F^{\prime\prime}_t}{\partial z_t}=\frac{(1+z_t)\left[4H_2(z_t)\frac{dH_2(z_t)}{dz_t}(1+j_t)-18H_0^2\Omega_m(1+z_t)^2\right]-2\left[2H_2^2(z_t)(1+j_t)-6H_0^2\Omega_m(1+z_t)^3\right]}{H_0^2(1-\Omega_m)(1+z_t)^3}\,,\\
        &\frac{\partial F^{\prime\prime}_t}{\partial j_t}=\frac{2H_2^2(z_t)}{H_0^2(1-\Omega_m)(1+z_t)^2}\,.
    \end{align}
\end{subequations}
In Eqs.~\eqref{36}--\eqref{38}, the derivative of the B\'ezier-interpolated $H_{2,t}$ with respect to $z_t$ are given by
\begin{equation}
    \frac{dH_{2,t}}{dz_t}=100\left[-\frac{2\alpha_0}{z_O}\left(1-\frac{z_t}{z_O}\right)+\frac{2\alpha_1}{z_O}\left(1-\frac{z_t}{z_O}\right)-\frac{2\alpha_1}{z_O}\left(\frac{z_t}{z_O}\right)+\frac{2\alpha_2}{z_O}\left(\frac{z_t}{z_O}\right)\right]\,.
\end{equation}

Now, concerning the error propagation for $\Omega_m$, we have
\begin{equation}
\delta\Omega_m={\Bigg\lvert\frac{\partial\Omega_m}{\partial z_t}\Bigg\lvert}\delta z_t +{\Bigg\lvert\frac{\partial\Omega_m}{\partial F_t}\Bigg\lvert}\delta F_t + {\Bigg\lvert\frac{\partial\Omega_m}{\partial F^\prime_t}\Bigg\lvert}\delta F^\prime_t\,,
\end{equation}
where
\begin{align}
    {\frac{\partial\Omega_m}{\partial z_t}} &=
-\frac{F^\prime_t}{2F_t-(1+z_t)(F^\prime_t-(1+z_t)^2)} +\frac{[2F_t-F^\prime_t(1+z_t)][F^\prime_t-3(1+z_t)^2]}{[2F_t-(1+z_t)(F^\prime_t-(1+z_t)^2)]^2}\,,\\
 {\frac{\partial\Omega_m}{\partial F_t}} &=
\frac{2}{2F_t-(1+z_t)(F^\prime_t-(1+z_t)^2)}
- \frac{2[2F_t-F^\prime_t(1+z_t)]}{[2F_t-(1+z_t)(F^\prime_t-(1+z_t)^2)]^2}\,,\\
{\frac{\partial\Omega_m}{\partial F^\prime_t}}&= -\frac{1+z_t}{2F_t-(1+z_t)(F^\prime_t-(1+z_t)^2)} + \frac{(1+z_t)[2F_t-F^\prime_t(1+z_t)]}{[2F_t-(1+z_t)(F^\prime_t-(1+z_t)^2)]^2}
\end{align}

\newpage
\twocolumngrid
%



\begin{thebibliography}{67}%
\makeatletter
\providecommand \@ifxundefined [1]{%
 \@ifx{#1\undefined}
}%
\providecommand \@ifnum [1]{%
 \ifnum #1\expandafter \@firstoftwo
 \else \expandafter \@secondoftwo
 \fi
}%
\providecommand \@ifx [1]{%
 \ifx #1\expandafter \@firstoftwo
 \else \expandafter \@secondoftwo
 \fi
}%
\providecommand \natexlab [1]{#1}%
\providecommand \enquote  [1]{``#1''}%
\providecommand \bibnamefont  [1]{#1}%
\providecommand \bibfnamefont [1]{#1}%
\providecommand \citenamefont [1]{#1}%
\providecommand \href@noop [0]{\@secondoftwo}%
\providecommand \href [0]{\begingroup \@sanitize@url \@href}%
\providecommand \@href[1]{\@@startlink{#1}\@@href}%
\providecommand \@@href[1]{\endgroup#1\@@endlink}%
\providecommand \@sanitize@url [0]{\catcode `\\12\catcode `\$12\catcode
  `\&12\catcode `\#12\catcode `\^12\catcode `\_12\catcode `\%12\relax}%
\providecommand \@@startlink[1]{}%
\providecommand \@@endlink[0]{}%
\providecommand \url  [0]{\begingroup\@sanitize@url \@url }%
\providecommand \@url [1]{\endgroup\@href {#1}{\urlprefix }}%
\providecommand \urlprefix  [0]{URL }%
\providecommand \Eprint [0]{\href }%
\providecommand \doibase [0]{http://dx.doi.org/}%
\providecommand \selectlanguage [0]{\@gobble}%
\providecommand \bibinfo  [0]{\@secondoftwo}%
\providecommand \bibfield  [0]{\@secondoftwo}%
\providecommand \translation [1]{[#1]}%
\providecommand \BibitemOpen [0]{}%
\providecommand \bibitemStop [0]{}%
\providecommand \bibitemNoStop [0]{.\EOS\space}%
\providecommand \EOS [0]{\spacefactor3000\relax}%
\providecommand \BibitemShut  [1]{\csname bibitem#1\endcsname}%
\let\auto@bib@innerbib\@empty
\bibitem [{\citenamefont {{Riess}}\ \emph {et~al.}(2001)\citenamefont
  {{Riess}}, \citenamefont {{Nugent}}, \citenamefont {{Gilliland}},
  \citenamefont {{Schmidt}}, \citenamefont {{Tonry}} \emph
  {et~al.}}]{2001ApJ...560...49R}%
  \BibitemOpen
  \bibfield  {author} {\bibinfo {author} {\bibfnamefont {A.~G.}\ \bibnamefont
  {{Riess}}}, \bibinfo {author} {\bibfnamefont {P.~E.}\ \bibnamefont
  {{Nugent}}}, \bibinfo {author} {\bibfnamefont {R.~L.}\ \bibnamefont
  {{Gilliland}}}, \bibinfo {author} {\bibfnamefont {B.~P.}\ \bibnamefont
  {{Schmidt}}}, \bibinfo {author} {\bibfnamefont {J.}~\bibnamefont {{Tonry}}},
  \emph {et~al.},\ }\href {\doibase 10.1086/322348} {\bibfield  {journal}
  {\bibinfo  {journal} {\apj}\ }\textbf {\bibinfo {volume} {560}},\ \bibinfo
  {pages} {49} (\bibinfo {year} {2001})},\ \Eprint
  {http://arxiv.org/abs/astro-ph/0104455} {arXiv:astro-ph/0104455 [astro-ph]}
  \BibitemShut {NoStop}%
\bibitem [{\citenamefont {{Guzzo}}\ \emph {et~al.}(2008)\citenamefont
  {{Guzzo}}, \citenamefont {{Pierleoni}}, \citenamefont {{Meneux}},
  \citenamefont {{Branchini}}, \citenamefont {{Le F{\`e}vre}} \emph
  {et~al.}}]{2008Natur.451..541G}%
  \BibitemOpen
  \bibfield  {author} {\bibinfo {author} {\bibfnamefont {L.}~\bibnamefont
  {{Guzzo}}}, \bibinfo {author} {\bibfnamefont {M.}~\bibnamefont
  {{Pierleoni}}}, \bibinfo {author} {\bibfnamefont {B.}~\bibnamefont
  {{Meneux}}}, \bibinfo {author} {\bibfnamefont {E.}~\bibnamefont
  {{Branchini}}}, \bibinfo {author} {\bibfnamefont {O.}~\bibnamefont {{Le
  F{\`e}vre}}},  \emph {et~al.},\ }\href {\doibase 10.1038/nature06555}
  {\bibfield  {journal} {\bibinfo  {journal} {\nat}\ }\textbf {\bibinfo
  {volume} {451}},\ \bibinfo {pages} {541} (\bibinfo {year} {2008})},\ \Eprint
  {http://arxiv.org/abs/0802.1944} {arXiv:0802.1944 [astro-ph]} \BibitemShut
  {NoStop}%
\bibitem [{\citenamefont {{Weinberg}}\ \emph {et~al.}(2013)\citenamefont
  {{Weinberg}}, \citenamefont {{Mortonson}}, \citenamefont {{Eisenstein}},
  \citenamefont {{Hirata}}, \citenamefont {{Riess}},\ and\ \citenamefont
  {{Rozo}}}]{2013PhR...530...87W}%
  \BibitemOpen
  \bibfield  {author} {\bibinfo {author} {\bibfnamefont {D.~H.}\ \bibnamefont
  {{Weinberg}}}, \bibinfo {author} {\bibfnamefont {M.~J.}\ \bibnamefont
  {{Mortonson}}}, \bibinfo {author} {\bibfnamefont {D.~J.}\ \bibnamefont
  {{Eisenstein}}}, \bibinfo {author} {\bibfnamefont {C.}~\bibnamefont
  {{Hirata}}}, \bibinfo {author} {\bibfnamefont {A.~G.}\ \bibnamefont
  {{Riess}}}, \ and\ \bibinfo {author} {\bibfnamefont {E.}~\bibnamefont
  {{Rozo}}},\ }\href {\doibase 10.1016/j.physrep.2013.05.001} {\bibfield
  {journal} {\bibinfo  {journal} {\physrep}\ }\textbf {\bibinfo {volume}
  {530}},\ \bibinfo {pages} {87} (\bibinfo {year} {2013})},\ \Eprint
  {http://arxiv.org/abs/1201.2434} {arXiv:1201.2434 [astro-ph.CO]} \BibitemShut
  {NoStop}%
\bibitem [{\citenamefont {{Riess}}\ \emph {et~al.}(1998)\citenamefont
  {{Riess}}, \citenamefont {{Filippenko}}, \citenamefont {{Challis}},
  \citenamefont {{Clocchiatti}}, \citenamefont {{Diercks}} \emph
  {et~al.}}]{riess}%
  \BibitemOpen
  \bibfield  {author} {\bibinfo {author} {\bibfnamefont {A.~G.}\ \bibnamefont
  {{Riess}}}, \bibinfo {author} {\bibfnamefont {A.~V.}\ \bibnamefont
  {{Filippenko}}}, \bibinfo {author} {\bibfnamefont {P.}~\bibnamefont
  {{Challis}}}, \bibinfo {author} {\bibfnamefont {A.}~\bibnamefont
  {{Clocchiatti}}}, \bibinfo {author} {\bibfnamefont {A.}~\bibnamefont
  {{Diercks}}},  \emph {et~al.},\ }\href {\doibase 10.1086/300499} {\bibfield
  {journal} {\bibinfo  {journal} {\aj}\ }\textbf {\bibinfo {volume} {116}},\
  \bibinfo {pages} {1009} (\bibinfo {year} {1998})},\ \Eprint
  {http://arxiv.org/abs/astro-ph/9805201} {arXiv:astro-ph/9805201 [astro-ph]}
  \BibitemShut {NoStop}%
\bibitem [{\citenamefont {{Perlmutter}}\ \emph {et~al.}(1999)\citenamefont
  {{Perlmutter}}, \citenamefont {{Aldering}}, \citenamefont {{Goldhaber}},
  \citenamefont {{Knop}}, \citenamefont {{Nugent}} \emph
  {et~al.}}]{Perlmutter}%
  \BibitemOpen
  \bibfield  {author} {\bibinfo {author} {\bibfnamefont {S.}~\bibnamefont
  {{Perlmutter}}}, \bibinfo {author} {\bibfnamefont {G.}~\bibnamefont
  {{Aldering}}}, \bibinfo {author} {\bibfnamefont {G.}~\bibnamefont
  {{Goldhaber}}}, \bibinfo {author} {\bibfnamefont {R.~A.}\ \bibnamefont
  {{Knop}}}, \bibinfo {author} {\bibfnamefont {P.}~\bibnamefont {{Nugent}}},
  \emph {et~al.},\ }\href {\doibase 10.1086/307221} {\bibfield  {journal}
  {\bibinfo  {journal} {\apj}\ }\textbf {\bibinfo {volume} {517}},\ \bibinfo
  {pages} {565} (\bibinfo {year} {1999})},\ \Eprint
  {http://arxiv.org/abs/astro-ph/9812133} {arXiv:astro-ph/9812133 [astro-ph]}
  \BibitemShut {NoStop}%
\bibitem [{\citenamefont {{Carroll}}(2001)}]{2001LRR.....4....1C}%
  \BibitemOpen
  \bibfield  {author} {\bibinfo {author} {\bibfnamefont {S.~M.}\ \bibnamefont
  {{Carroll}}},\ }\href {\doibase 10.12942/lrr-2001-1} {\bibfield  {journal}
  {\bibinfo  {journal} {Living Reviews in Relativity}\ }\textbf {\bibinfo
  {volume} {4}},\ \bibinfo {eid} {1} (\bibinfo {year} {2001})},\ \Eprint
  {http://arxiv.org/abs/astro-ph/0004075} {arXiv:astro-ph/0004075 [astro-ph]}
  \BibitemShut {NoStop}%
\bibitem [{\citenamefont {{Weinberg}}(2000)}]{2000astro.ph..5265W}%
  \BibitemOpen
  \bibfield  {author} {\bibinfo {author} {\bibfnamefont {S.}~\bibnamefont
  {{Weinberg}}},\ }\href {\doibase 10.48550/arXiv.astro-ph/0005265} {\bibfield
  {journal} {\bibinfo  {journal} {arXiv e-prints}\ ,\ \bibinfo {eid}
  {astro-ph/0005265}} (\bibinfo {year} {2000})},\ \Eprint
  {http://arxiv.org/abs/astro-ph/0005265} {arXiv:astro-ph/0005265 [astro-ph]}
  \BibitemShut {NoStop}%
\bibitem [{\citenamefont {{Peebles}}\ and\ \citenamefont
  {{Ratra}}(2003)}]{2003RvMP...75..559P}%
  \BibitemOpen
  \bibfield  {author} {\bibinfo {author} {\bibfnamefont {P.~J.}\ \bibnamefont
  {{Peebles}}}\ and\ \bibinfo {author} {\bibfnamefont {B.}~\bibnamefont
  {{Ratra}}},\ }\href {\doibase 10.1103/RevModPhys.75.559} {\bibfield
  {journal} {\bibinfo  {journal} {Reviews of Modern Physics}\ }\textbf
  {\bibinfo {volume} {75}},\ \bibinfo {pages} {559} (\bibinfo {year} {2003})},\
  \Eprint {http://arxiv.org/abs/astro-ph/0207347} {arXiv:astro-ph/0207347
  [astro-ph]} \BibitemShut {NoStop}%
\bibitem [{\citenamefont {{Padmanabhan}}(2003)}]{2003PhR...380..235P}%
  \BibitemOpen
  \bibfield  {author} {\bibinfo {author} {\bibfnamefont {T.}~\bibnamefont
  {{Padmanabhan}}},\ }\href {\doibase 10.1016/S0370-1573(03)00120-0} {\bibfield
   {journal} {\bibinfo  {journal} {\physrep}\ }\textbf {\bibinfo {volume}
  {380}},\ \bibinfo {pages} {235} (\bibinfo {year} {2003})},\ \Eprint
  {http://arxiv.org/abs/hep-th/0212290} {arXiv:hep-th/0212290 [hep-th]}
  \BibitemShut {NoStop}%
\bibitem [{\citenamefont {{Copeland}}\ \emph {et~al.}(2006)\citenamefont
  {{Copeland}}, \citenamefont {{Sami}},\ and\ \citenamefont
  {{Tsujikawa}}}]{copeland}%
  \BibitemOpen
  \bibfield  {author} {\bibinfo {author} {\bibfnamefont {E.~J.}\ \bibnamefont
  {{Copeland}}}, \bibinfo {author} {\bibfnamefont {M.}~\bibnamefont {{Sami}}},
  \ and\ \bibinfo {author} {\bibfnamefont {S.}~\bibnamefont {{Tsujikawa}}},\
  }\href {\doibase 10.1142/S021827180600942X} {\bibfield  {journal} {\bibinfo
  {journal} {International Journal of Modern Physics D}\ }\textbf {\bibinfo
  {volume} {15}},\ \bibinfo {pages} {1753} (\bibinfo {year} {2006})},\ \Eprint
  {http://arxiv.org/abs/hep-th/0603057} {arXiv:hep-th/0603057 [hep-th]}
  \BibitemShut {NoStop}%
\bibitem [{\citenamefont {{Capozziello}}\ \emph {et~al.}(2013)\citenamefont
  {{Capozziello}}, \citenamefont {{De Laurentis}}, \citenamefont {{Luongo}},\
  and\ \citenamefont {{Ruggeri}}}]{capozziello2013}%
  \BibitemOpen
  \bibfield  {author} {\bibinfo {author} {\bibfnamefont {S.}~\bibnamefont
  {{Capozziello}}}, \bibinfo {author} {\bibfnamefont {M.}~\bibnamefont {{De
  Laurentis}}}, \bibinfo {author} {\bibfnamefont {O.}~\bibnamefont {{Luongo}}},
  \ and\ \bibinfo {author} {\bibfnamefont {A.}~\bibnamefont {{Ruggeri}}},\
  }\href {\doibase 10.3390/galaxies1030216} {\bibfield  {journal} {\bibinfo
  {journal} {Galaxies}\ }\textbf {\bibinfo {volume} {1}},\ \bibinfo {pages}
  {216} (\bibinfo {year} {2013})},\ \Eprint {http://arxiv.org/abs/1312.1825}
  {arXiv:1312.1825 [gr-qc]} \BibitemShut {NoStop}%
\bibitem [{\citenamefont {{Sahni}}(2002)}]{2002CQGra..19.3435S}%
  \BibitemOpen
  \bibfield  {author} {\bibinfo {author} {\bibfnamefont {V.}~\bibnamefont
  {{Sahni}}},\ }\href {\doibase 10.1088/0264-9381/19/13/304} {\bibfield
  {journal} {\bibinfo  {journal} {Classical and Quantum Gravity}\ }\textbf
  {\bibinfo {volume} {19}},\ \bibinfo {pages} {3435} (\bibinfo {year}
  {2002})},\ \Eprint {http://arxiv.org/abs/astro-ph/0202076}
  {arXiv:astro-ph/0202076 [astro-ph]} \BibitemShut {NoStop}%
\bibitem [{\citenamefont {{Tsujikawa}}(2013)}]{2013CQGra..30u4003T}%
  \BibitemOpen
  \bibfield  {author} {\bibinfo {author} {\bibfnamefont {S.}~\bibnamefont
  {{Tsujikawa}}},\ }\href {\doibase 10.1088/0264-9381/30/21/214003} {\bibfield
  {journal} {\bibinfo  {journal} {Classical and Quantum Gravity}\ }\textbf
  {\bibinfo {volume} {30}},\ \bibinfo {eid} {214003} (\bibinfo {year}
  {2013})},\ \Eprint {http://arxiv.org/abs/1304.1961} {arXiv:1304.1961 [gr-qc]}
  \BibitemShut {NoStop}%
\bibitem [{\citenamefont {{Chevallier}}\ and\ \citenamefont
  {{Polarski}}(2001)}]{2001IJMPD..10..213C}%
  \BibitemOpen
  \bibfield  {author} {\bibinfo {author} {\bibfnamefont {M.}~\bibnamefont
  {{Chevallier}}}\ and\ \bibinfo {author} {\bibfnamefont {D.}~\bibnamefont
  {{Polarski}}},\ }\href {\doibase 10.1142/S0218271801000822} {\bibfield
  {journal} {\bibinfo  {journal} {International Journal of Modern Physics D}\
  }\textbf {\bibinfo {volume} {10}},\ \bibinfo {pages} {213} (\bibinfo {year}
  {2001})},\ \Eprint {http://arxiv.org/abs/gr-qc/0009008} {arXiv:gr-qc/0009008
  [gr-qc]} \BibitemShut {NoStop}%
\bibitem [{\citenamefont {{Linder}}(2003)}]{2003PhRvL..90i1301L}%
  \BibitemOpen
  \bibfield  {author} {\bibinfo {author} {\bibfnamefont {E.~V.}\ \bibnamefont
  {{Linder}}},\ }\href {\doibase 10.1103/PhysRevLett.90.091301} {\bibfield
  {journal} {\bibinfo  {journal} {\prl}\ }\textbf {\bibinfo {volume} {90}},\
  \bibinfo {eid} {091301} (\bibinfo {year} {2003})},\ \Eprint
  {http://arxiv.org/abs/astro-ph/0208512} {arXiv:astro-ph/0208512 [astro-ph]}
  \BibitemShut {NoStop}%
\bibitem [{\citenamefont {{Corasaniti}}\ and\ \citenamefont
  {{Copeland}}(2003)}]{2003PhRvD..67f3521C}%
  \BibitemOpen
  \bibfield  {author} {\bibinfo {author} {\bibfnamefont {P.~S.}\ \bibnamefont
  {{Corasaniti}}}\ and\ \bibinfo {author} {\bibfnamefont {E.~J.}\ \bibnamefont
  {{Copeland}}},\ }\href {\doibase 10.1103/PhysRevD.67.063521} {\bibfield
  {journal} {\bibinfo  {journal} {\prd}\ }\textbf {\bibinfo {volume} {67}},\
  \bibinfo {eid} {063521} (\bibinfo {year} {2003})},\ \Eprint
  {http://arxiv.org/abs/astro-ph/0205544} {arXiv:astro-ph/0205544 [astro-ph]}
  \BibitemShut {NoStop}%
\bibitem [{\citenamefont {{Shafieloo}}(2007)}]{2007MNRAS.380.1573S}%
  \BibitemOpen
  \bibfield  {author} {\bibinfo {author} {\bibfnamefont {A.}~\bibnamefont
  {{Shafieloo}}},\ }\href {\doibase 10.1111/j.1365-2966.2007.12175.x}
  {\bibfield  {journal} {\bibinfo  {journal} {\mnras}\ }\textbf {\bibinfo
  {volume} {380}},\ \bibinfo {pages} {1573} (\bibinfo {year} {2007})},\ \Eprint
  {http://arxiv.org/abs/astro-ph/0703034} {arXiv:astro-ph/0703034 [astro-ph]}
  \BibitemShut {NoStop}%
\bibitem [{\citenamefont {{Nesseris}}\ and\ \citenamefont
  {{Garc{\'\i}a-Bellido}}(2013)}]{2013PhRvD..88f3521N}%
  \BibitemOpen
  \bibfield  {author} {\bibinfo {author} {\bibfnamefont {S.}~\bibnamefont
  {{Nesseris}}}\ and\ \bibinfo {author} {\bibfnamefont {J.}~\bibnamefont
  {{Garc{\'\i}a-Bellido}}},\ }\href {\doibase 10.1103/PhysRevD.88.063521}
  {\bibfield  {journal} {\bibinfo  {journal} {\prd}\ }\textbf {\bibinfo
  {volume} {88}},\ \bibinfo {eid} {063521} (\bibinfo {year} {2013})},\ \Eprint
  {http://arxiv.org/abs/1306.4885} {arXiv:1306.4885 [astro-ph.CO]} \BibitemShut
  {NoStop}%
\bibitem [{\citenamefont {{Sahni}}\ \emph {et~al.}(2014)\citenamefont
  {{Sahni}}, \citenamefont {{Shafieloo}},\ and\ \citenamefont
  {{Starobinsky}}}]{2014ApJ...793L..40S}%
  \BibitemOpen
  \bibfield  {author} {\bibinfo {author} {\bibfnamefont {V.}~\bibnamefont
  {{Sahni}}}, \bibinfo {author} {\bibfnamefont {A.}~\bibnamefont
  {{Shafieloo}}}, \ and\ \bibinfo {author} {\bibfnamefont {A.~A.}\ \bibnamefont
  {{Starobinsky}}},\ }\href {\doibase 10.1088/2041-8205/793/2/L40} {\bibfield
  {journal} {\bibinfo  {journal} {\apjl}\ }\textbf {\bibinfo {volume} {793}},\
  \bibinfo {eid} {L40} (\bibinfo {year} {2014})},\ \Eprint
  {http://arxiv.org/abs/1406.2209} {arXiv:1406.2209 [astro-ph.CO]} \BibitemShut
  {NoStop}%
\bibitem [{\citenamefont {{Harrison}}(1976)}]{Harrison}%
  \BibitemOpen
  \bibfield  {author} {\bibinfo {author} {\bibfnamefont {E.~R.}\ \bibnamefont
  {{Harrison}}},\ }\href {\doibase 10.1038/260591a0} {\bibfield  {journal}
  {\bibinfo  {journal} {\nat}\ }\textbf {\bibinfo {volume} {260}},\ \bibinfo
  {pages} {591} (\bibinfo {year} {1976})}\BibitemShut {NoStop}%
\bibitem [{\citenamefont {{Visser}}(2005)}]{Visser}%
  \BibitemOpen
  \bibfield  {author} {\bibinfo {author} {\bibfnamefont {M.}~\bibnamefont
  {{Visser}}},\ }\href {\doibase 10.1007/s10714-005-0134-8} {\bibfield
  {journal} {\bibinfo  {journal} {General Relativity and Gravitation}\ }\textbf
  {\bibinfo {volume} {37}},\ \bibinfo {pages} {1541} (\bibinfo {year}
  {2005})}\BibitemShut {NoStop}%
\bibitem [{\citenamefont {{Dunsby}}\ and\ \citenamefont
  {{Luongo}}(2016)}]{2016IJGMM..1330002D}%
  \BibitemOpen
  \bibfield  {author} {\bibinfo {author} {\bibfnamefont {P.~K.~S.}\
  \bibnamefont {{Dunsby}}}\ and\ \bibinfo {author} {\bibfnamefont
  {O.}~\bibnamefont {{Luongo}}},\ }\href {\doibase 10.1142/S0219887816300026}
  {\bibfield  {journal} {\bibinfo  {journal} {International Journal of
  Geometric Methods in Modern Physics}\ }\textbf {\bibinfo {volume} {13}},\
  \bibinfo {eid} {1630002-606} (\bibinfo {year} {2016})},\ \Eprint
  {http://arxiv.org/abs/1511.06532} {arXiv:1511.06532 [gr-qc]} \BibitemShut
  {NoStop}%
\bibitem [{\citenamefont {Capozziello}\ \emph {et~al.}(2011)\citenamefont
  {Capozziello}, \citenamefont {Lazkoz},\ and\ \citenamefont
  {Salzano}}]{salzano}%
  \BibitemOpen
  \bibfield  {author} {\bibinfo {author} {\bibfnamefont {S.}~\bibnamefont
  {Capozziello}}, \bibinfo {author} {\bibfnamefont {R.}~\bibnamefont {Lazkoz}},
  \ and\ \bibinfo {author} {\bibfnamefont {V.}~\bibnamefont {Salzano}},\ }\href
  {\doibase 10.1103/PhysRevD.84.124061} {\bibfield  {journal} {\bibinfo
  {journal} {Phys. Rev. D}\ }\textbf {\bibinfo {volume} {84}},\ \bibinfo
  {pages} {124061} (\bibinfo {year} {2011})},\ \Eprint
  {http://arxiv.org/abs/1104.3096} {arXiv:1104.3096 [astro-ph.CO]} \BibitemShut
  {NoStop}%
\bibitem [{\citenamefont {{Catto{\"e}n}}\ and\ \citenamefont
  {{Visser}}(2007)}]{2007CQGra..24.5985C}%
  \BibitemOpen
  \bibfield  {author} {\bibinfo {author} {\bibfnamefont {C.}~\bibnamefont
  {{Catto{\"e}n}}}\ and\ \bibinfo {author} {\bibfnamefont {M.}~\bibnamefont
  {{Visser}}},\ }\href {\doibase 10.1088/0264-9381/24/23/018} {\bibfield
  {journal} {\bibinfo  {journal} {Classical and Quantum Gravity}\ }\textbf
  {\bibinfo {volume} {24}},\ \bibinfo {pages} {5985} (\bibinfo {year}
  {2007})},\ \Eprint {http://arxiv.org/abs/0710.1887} {arXiv:0710.1887 [gr-qc]}
  \BibitemShut {NoStop}%
\bibitem [{\citenamefont {{Catto{\"e}n}}\ and\ \citenamefont
  {{Visser}}(2008)}]{2008PhRvD..78f3501C}%
  \BibitemOpen
  \bibfield  {author} {\bibinfo {author} {\bibfnamefont {C.}~\bibnamefont
  {{Catto{\"e}n}}}\ and\ \bibinfo {author} {\bibfnamefont {M.}~\bibnamefont
  {{Visser}}},\ }\href {\doibase 10.1103/PhysRevD.78.063501} {\bibfield
  {journal} {\bibinfo  {journal} {\prd}\ }\textbf {\bibinfo {volume} {78}},\
  \bibinfo {eid} {063501} (\bibinfo {year} {2008})},\ \Eprint
  {http://arxiv.org/abs/0809.0537} {arXiv:0809.0537 [gr-qc]} \BibitemShut
  {NoStop}%
\bibitem [{\citenamefont {{Aviles}}\ \emph {et~al.}(2012)\citenamefont
  {{Aviles}}, \citenamefont {{Gruber}}, \citenamefont {{Luongo}},\ and\
  \citenamefont {{Quevedo}}}]{2012PhRvD..86l3516A}%
  \BibitemOpen
  \bibfield  {author} {\bibinfo {author} {\bibfnamefont {A.}~\bibnamefont
  {{Aviles}}}, \bibinfo {author} {\bibfnamefont {C.}~\bibnamefont {{Gruber}}},
  \bibinfo {author} {\bibfnamefont {O.}~\bibnamefont {{Luongo}}}, \ and\
  \bibinfo {author} {\bibfnamefont {H.}~\bibnamefont {{Quevedo}}},\ }\href
  {\doibase 10.1103/PhysRevD.86.123516} {\bibfield  {journal} {\bibinfo
  {journal} {\prd}\ }\textbf {\bibinfo {volume} {86}},\ \bibinfo {eid} {123516}
  (\bibinfo {year} {2012})},\ \Eprint {http://arxiv.org/abs/1204.2007}
  {arXiv:1204.2007 [astro-ph.CO]} \BibitemShut {NoStop}%
\bibitem [{\citenamefont {{Gruber}}\ and\ \citenamefont
  {{Luongo}}(2014)}]{2014PhRvD..89j3506G}%
  \BibitemOpen
  \bibfield  {author} {\bibinfo {author} {\bibfnamefont {C.}~\bibnamefont
  {{Gruber}}}\ and\ \bibinfo {author} {\bibfnamefont {O.}~\bibnamefont
  {{Luongo}}},\ }\href {\doibase 10.1103/PhysRevD.89.103506} {\bibfield
  {journal} {\bibinfo  {journal} {\prd}\ }\textbf {\bibinfo {volume} {89}},\
  \bibinfo {eid} {103506} (\bibinfo {year} {2014})},\ \Eprint
  {http://arxiv.org/abs/1309.3215} {arXiv:1309.3215 [gr-qc]} \BibitemShut
  {NoStop}%
\bibitem [{\citenamefont {{Capozziello}}\ \emph {et~al.}(2022)\citenamefont
  {{Capozziello}}, \citenamefont {{Dunsby}},\ and\ \citenamefont
  {{Luongo}}}]{capozziellodunsbyluongo}%
  \BibitemOpen
  \bibfield  {author} {\bibinfo {author} {\bibfnamefont {S.}~\bibnamefont
  {{Capozziello}}}, \bibinfo {author} {\bibfnamefont {P.~K.~S.}\ \bibnamefont
  {{Dunsby}}}, \ and\ \bibinfo {author} {\bibfnamefont {O.}~\bibnamefont
  {{Luongo}}},\ }\href {\doibase 10.1093/mnras/stab3187} {\bibfield  {journal}
  {\bibinfo  {journal} {\mnras}\ }\textbf {\bibinfo {volume} {509}},\ \bibinfo
  {pages} {5399} (\bibinfo {year} {2022})},\ \Eprint
  {http://arxiv.org/abs/2106.15579} {arXiv:2106.15579 [astro-ph.CO]}
  \BibitemShut {NoStop}%
\bibitem [{\citenamefont {{Ghirlanda}}\ \emph
  {et~al.}(2004{\natexlab{a}})\citenamefont {{Ghirlanda}}, \citenamefont
  {{Ghisellini}}, \citenamefont {{Lazzati}},\ and\ \citenamefont
  {{Firmani}}}]{ghirlanda1}%
  \BibitemOpen
  \bibfield  {author} {\bibinfo {author} {\bibfnamefont {G.}~\bibnamefont
  {{Ghirlanda}}}, \bibinfo {author} {\bibfnamefont {G.}~\bibnamefont
  {{Ghisellini}}}, \bibinfo {author} {\bibfnamefont {D.}~\bibnamefont
  {{Lazzati}}}, \ and\ \bibinfo {author} {\bibfnamefont {C.}~\bibnamefont
  {{Firmani}}},\ }\href {\doibase 10.1086/424915} {\bibfield  {journal}
  {\bibinfo  {journal} {\apjl}\ }\textbf {\bibinfo {volume} {613}},\ \bibinfo
  {pages} {L13} (\bibinfo {year} {2004}{\natexlab{a}})},\ \Eprint
  {http://arxiv.org/abs/astro-ph/0408350} {arXiv:astro-ph/0408350 [astro-ph]}
  \BibitemShut {NoStop}%
\bibitem [{\citenamefont {{Firmani}}\ \emph {et~al.}(2005)\citenamefont
  {{Firmani}}, \citenamefont {{Ghisellini}}, \citenamefont {{Ghirlanda}},\ and\
  \citenamefont {{Avila-Reese}}}]{firmani}%
  \BibitemOpen
  \bibfield  {author} {\bibinfo {author} {\bibfnamefont {C.}~\bibnamefont
  {{Firmani}}}, \bibinfo {author} {\bibfnamefont {G.}~\bibnamefont
  {{Ghisellini}}}, \bibinfo {author} {\bibfnamefont {G.}~\bibnamefont
  {{Ghirlanda}}}, \ and\ \bibinfo {author} {\bibfnamefont {V.}~\bibnamefont
  {{Avila-Reese}}},\ }\href {\doibase 10.1111/j.1745-3933.2005.00023.x}
  {\bibfield  {journal} {\bibinfo  {journal} {\mnras}\ }\textbf {\bibinfo
  {volume} {360}},\ \bibinfo {pages} {L1} (\bibinfo {year} {2005})},\ \Eprint
  {http://arxiv.org/abs/astro-ph/0501395} {arXiv:astro-ph/0501395 [astro-ph]}
  \BibitemShut {NoStop}%
\bibitem [{\citenamefont {{Luongo}}\ and\ \citenamefont
  {{Muccino}}(2021{\natexlab{a}})}]{luongomuccino2021}%
  \BibitemOpen
  \bibfield  {author} {\bibinfo {author} {\bibfnamefont {O.}~\bibnamefont
  {{Luongo}}}\ and\ \bibinfo {author} {\bibfnamefont {M.}~\bibnamefont
  {{Muccino}}},\ }\href {\doibase 10.3390/galaxies9040077} {\bibfield
  {journal} {\bibinfo  {journal} {Galaxies}\ }\textbf {\bibinfo {volume} {9}},\
  \bibinfo {pages} {77} (\bibinfo {year} {2021}{\natexlab{a}})},\ \Eprint
  {http://arxiv.org/abs/2110.14408} {arXiv:2110.14408 [astro-ph.HE]}
  \BibitemShut {NoStop}%
\bibitem [{\citenamefont {Demianski}\ \emph {et~al.}(2021)\citenamefont
  {Demianski}, \citenamefont {Piedipalumbo}, \citenamefont {Sawant},\ and\
  \citenamefont {Amati}}]{ester1}%
  \BibitemOpen
  \bibfield  {author} {\bibinfo {author} {\bibfnamefont {M.}~\bibnamefont
  {Demianski}}, \bibinfo {author} {\bibfnamefont {E.}~\bibnamefont
  {Piedipalumbo}}, \bibinfo {author} {\bibfnamefont {D.}~\bibnamefont
  {Sawant}}, \ and\ \bibinfo {author} {\bibfnamefont {L.}~\bibnamefont
  {Amati}},\ }\href {\doibase 10.1093/mnras/stab1669} {\bibfield  {journal}
  {\bibinfo  {journal} {Mon. Not. Roy. Astron. Soc.}\ }\textbf {\bibinfo
  {volume} {506}},\ \bibinfo {pages} {903} (\bibinfo {year} {2021})},\ \Eprint
  {http://arxiv.org/abs/1911.08228} {arXiv:1911.08228 [astro-ph.CO]}
  \BibitemShut {NoStop}%
\bibitem [{\citenamefont {Lusso}\ \emph {et~al.}(2019)\citenamefont {Lusso},
  \citenamefont {Piedipalumbo}, \citenamefont {Risaliti}, \citenamefont
  {Paolillo}, \citenamefont {Bisogni} \emph {et~al.}}]{ester2}%
  \BibitemOpen
  \bibfield  {author} {\bibinfo {author} {\bibfnamefont {E.}~\bibnamefont
  {Lusso}}, \bibinfo {author} {\bibfnamefont {E.}~\bibnamefont {Piedipalumbo}},
  \bibinfo {author} {\bibfnamefont {G.}~\bibnamefont {Risaliti}}, \bibinfo
  {author} {\bibfnamefont {M.}~\bibnamefont {Paolillo}}, \bibinfo {author}
  {\bibfnamefont {S.}~\bibnamefont {Bisogni}},  \emph {et~al.},\ }\href
  {\doibase 10.1051/0004-6361/201936223} {\bibfield  {journal} {\bibinfo
  {journal} {Astron. Astrophys.}\ }\textbf {\bibinfo {volume} {628}},\ \bibinfo
  {pages} {L4} (\bibinfo {year} {2019})},\ \Eprint
  {http://arxiv.org/abs/1907.07692} {arXiv:1907.07692 [astro-ph.CO]}
  \BibitemShut {NoStop}%
\bibitem [{\citenamefont {Piedipalumbo}\ \emph {et~al.}(2014)\citenamefont
  {Piedipalumbo}, \citenamefont {Della~Moglie}, \citenamefont {De~Laurentis},\
  and\ \citenamefont {Scudellaro}}]{ester3}%
  \BibitemOpen
  \bibfield  {author} {\bibinfo {author} {\bibfnamefont {E.}~\bibnamefont
  {Piedipalumbo}}, \bibinfo {author} {\bibfnamefont {E.}~\bibnamefont
  {Della~Moglie}}, \bibinfo {author} {\bibfnamefont {M.}~\bibnamefont
  {De~Laurentis}}, \ and\ \bibinfo {author} {\bibfnamefont {P.}~\bibnamefont
  {Scudellaro}},\ }\href {\doibase 10.1093/mnras/stu790} {\bibfield  {journal}
  {\bibinfo  {journal} {Mon. Not. Roy. Astron. Soc.}\ }\textbf {\bibinfo
  {volume} {441}},\ \bibinfo {pages} {3643} (\bibinfo {year} {2014})},\ \Eprint
  {http://arxiv.org/abs/1311.0995} {arXiv:1311.0995 [astro-ph.CO]} \BibitemShut
  {NoStop}%
\bibitem [{\citenamefont {Bargiacchi}\ \emph {et~al.}(2023)\citenamefont
  {Bargiacchi}, \citenamefont {Dainotti},\ and\ \citenamefont
  {Capozziello}}]{giada1}%
  \BibitemOpen
  \bibfield  {author} {\bibinfo {author} {\bibfnamefont {G.}~\bibnamefont
  {Bargiacchi}}, \bibinfo {author} {\bibfnamefont {M.~G.}\ \bibnamefont
  {Dainotti}}, \ and\ \bibinfo {author} {\bibfnamefont {S.}~\bibnamefont
  {Capozziello}},\ }\href {\doibase 10.1093/mnras/stad2326} {\bibfield
  {journal} {\bibinfo  {journal} {Mon. Not. Roy. Astron. Soc.}\ }\textbf
  {\bibinfo {volume} {525}},\ \bibinfo {pages} {3104} (\bibinfo {year}
  {2023})},\ \Eprint {http://arxiv.org/abs/2307.15359} {arXiv:2307.15359
  [astro-ph.CO]} \BibitemShut {NoStop}%
\bibitem [{\citenamefont {Dainotti}\ \emph {et~al.}(2023)\citenamefont
  {Dainotti}, \citenamefont {Bargiacchi}, \citenamefont {Lenart}, \citenamefont
  {Nagataki},\ and\ \citenamefont {Capozziello}}]{giada2}%
  \BibitemOpen
  \bibfield  {author} {\bibinfo {author} {\bibfnamefont {M.~G.}\ \bibnamefont
  {Dainotti}}, \bibinfo {author} {\bibfnamefont {G.}~\bibnamefont
  {Bargiacchi}}, \bibinfo {author} {\bibfnamefont {A.~L.}\ \bibnamefont
  {Lenart}}, \bibinfo {author} {\bibfnamefont {S.}~\bibnamefont {Nagataki}}, \
  and\ \bibinfo {author} {\bibfnamefont {S.}~\bibnamefont {Capozziello}},\
  }\href {\doibase 10.3847/1538-4357/accea0} {\bibfield  {journal} {\bibinfo
  {journal} {Astrophys. J.}\ }\textbf {\bibinfo {volume} {950}},\ \bibinfo
  {pages} {45} (\bibinfo {year} {2023})},\ \Eprint
  {http://arxiv.org/abs/2305.19668} {arXiv:2305.19668 [astro-ph.CO]}
  \BibitemShut {NoStop}%
\bibitem [{\citenamefont {{Amati}}\ \emph {et~al.}(2019)\citenamefont
  {{Amati}}, \citenamefont {{D'Agostino}}, \citenamefont {{Luongo}},
  \citenamefont {{Muccino}},\ and\ \citenamefont {{Tantalo}}}]{bezier}%
  \BibitemOpen
  \bibfield  {author} {\bibinfo {author} {\bibfnamefont {L.}~\bibnamefont
  {{Amati}}}, \bibinfo {author} {\bibfnamefont {R.}~\bibnamefont
  {{D'Agostino}}}, \bibinfo {author} {\bibfnamefont {O.}~\bibnamefont
  {{Luongo}}}, \bibinfo {author} {\bibfnamefont {M.}~\bibnamefont {{Muccino}}},
  \ and\ \bibinfo {author} {\bibfnamefont {M.}~\bibnamefont {{Tantalo}}},\
  }\href {\doibase 10.1093/mnrasl/slz056} {\bibfield  {journal} {\bibinfo
  {journal} {\mnras}\ }\textbf {\bibinfo {volume} {486}},\ \bibinfo {pages}
  {L46} (\bibinfo {year} {2019})},\ \Eprint {http://arxiv.org/abs/1811.08934}
  {arXiv:1811.08934 [astro-ph.HE]} \BibitemShut {NoStop}%
\bibitem [{\citenamefont {{Luongo}}\ and\ \citenamefont
  {{Muccino}}(2021{\natexlab{b}})}]{2021MNRAS.503.4581L}%
  \BibitemOpen
  \bibfield  {author} {\bibinfo {author} {\bibfnamefont {O.}~\bibnamefont
  {{Luongo}}}\ and\ \bibinfo {author} {\bibfnamefont {M.}~\bibnamefont
  {{Muccino}}},\ }\href {\doibase 10.1093/mnras/stab795} {\bibfield  {journal}
  {\bibinfo  {journal} {\mnras}\ }\textbf {\bibinfo {volume} {503}},\ \bibinfo
  {pages} {4581} (\bibinfo {year} {2021}{\natexlab{b}})},\ \Eprint
  {http://arxiv.org/abs/2011.13590} {arXiv:2011.13590 [astro-ph.CO]}
  \BibitemShut {NoStop}%
\bibitem [{\citenamefont {{Luongo}}\ and\ \citenamefont
  {{Muccino}}(2023)}]{2023MNRAS.518.2247L}%
  \BibitemOpen
  \bibfield  {author} {\bibinfo {author} {\bibfnamefont {O.}~\bibnamefont
  {{Luongo}}}\ and\ \bibinfo {author} {\bibfnamefont {M.}~\bibnamefont
  {{Muccino}}},\ }\href {\doibase 10.1093/mnras/stac2925} {\bibfield  {journal}
  {\bibinfo  {journal} {\mnras}\ }\textbf {\bibinfo {volume} {518}},\ \bibinfo
  {pages} {2247} (\bibinfo {year} {2023})},\ \Eprint
  {http://arxiv.org/abs/2207.00440} {arXiv:2207.00440 [astro-ph.CO]}
  \BibitemShut {NoStop}%
\bibitem [{\citenamefont {{Alfano}}\ \emph {et~al.}(2023)\citenamefont
  {{Alfano}}, \citenamefont {{Luongo}},\ and\ \citenamefont
  {{Muccino}}}]{2023arXiv231105324A}%
  \BibitemOpen
  \bibfield  {author} {\bibinfo {author} {\bibfnamefont {A.~C.}\ \bibnamefont
  {{Alfano}}}, \bibinfo {author} {\bibfnamefont {O.}~\bibnamefont {{Luongo}}},
  \ and\ \bibinfo {author} {\bibfnamefont {M.}~\bibnamefont {{Muccino}}},\
  }\href {\doibase 10.48550/arXiv.2311.05324} {\bibfield  {journal} {\bibinfo
  {journal} {arXiv e-prints}\ ,\ \bibinfo {eid} {arXiv:2311.05324}} (\bibinfo
  {year} {2023})},\ \Eprint {http://arxiv.org/abs/2311.05324} {arXiv:2311.05324
  [astro-ph.CO]} \BibitemShut {NoStop}%
\bibitem [{\citenamefont {{Amati}}\ and\ \citenamefont {{Della
  Valle}}(2013)}]{amatiref}%
  \BibitemOpen
  \bibfield  {author} {\bibinfo {author} {\bibfnamefont {L.}~\bibnamefont
  {{Amati}}}\ and\ \bibinfo {author} {\bibfnamefont {M.}~\bibnamefont {{Della
  Valle}}},\ }\href {\doibase 10.1142/S0218271813300280} {\bibfield  {journal}
  {\bibinfo  {journal} {International Journal of Modern Physics D}\ }\textbf
  {\bibinfo {volume} {22}},\ \bibinfo {eid} {1330028} (\bibinfo {year}
  {2013})},\ \Eprint {http://arxiv.org/abs/1310.3141} {arXiv:1310.3141
  [astro-ph.CO]} \BibitemShut {NoStop}%
\bibitem [{\citenamefont {{Izzo}}\ \emph {et~al.}(2015)\citenamefont {{Izzo}},
  \citenamefont {{Muccino}}, \citenamefont {{Zaninoni}}, \citenamefont
  {{Amati}},\ and\ \citenamefont {{Della Valle}}}]{comboref}%
  \BibitemOpen
  \bibfield  {author} {\bibinfo {author} {\bibfnamefont {L.}~\bibnamefont
  {{Izzo}}}, \bibinfo {author} {\bibfnamefont {M.}~\bibnamefont {{Muccino}}},
  \bibinfo {author} {\bibfnamefont {E.}~\bibnamefont {{Zaninoni}}}, \bibinfo
  {author} {\bibfnamefont {L.}~\bibnamefont {{Amati}}}, \ and\ \bibinfo
  {author} {\bibfnamefont {M.}~\bibnamefont {{Della Valle}}},\ }\href {\doibase
  10.1051/0004-6361/201526461} {\bibfield  {journal} {\bibinfo  {journal}
  {\aap}\ }\textbf {\bibinfo {volume} {582}},\ \bibinfo {eid} {A115} (\bibinfo
  {year} {2015})},\ \Eprint {http://arxiv.org/abs/1508.05898} {arXiv:1508.05898
  [astro-ph.CO]} \BibitemShut {NoStop}%
\bibitem [{\citenamefont {{Yonetoku}}\ \emph {et~al.}(2004)\citenamefont
  {{Yonetoku}}, \citenamefont {{Murakami}}, \citenamefont {{Nakamura}},
  \citenamefont {{Yamazaki}}, \citenamefont {{Inoue}},\ and\ \citenamefont
  {{Ioka}}}]{Yonetoku}%
  \BibitemOpen
  \bibfield  {author} {\bibinfo {author} {\bibfnamefont {D.}~\bibnamefont
  {{Yonetoku}}}, \bibinfo {author} {\bibfnamefont {T.}~\bibnamefont
  {{Murakami}}}, \bibinfo {author} {\bibfnamefont {T.}~\bibnamefont
  {{Nakamura}}}, \bibinfo {author} {\bibfnamefont {R.}~\bibnamefont
  {{Yamazaki}}}, \bibinfo {author} {\bibfnamefont {A.~K.}\ \bibnamefont
  {{Inoue}}}, \ and\ \bibinfo {author} {\bibfnamefont {K.}~\bibnamefont
  {{Ioka}}},\ }\href {\doibase 10.1086/421285} {\bibfield  {journal} {\bibinfo
  {journal} {\apj}\ }\textbf {\bibinfo {volume} {609}},\ \bibinfo {pages} {935}
  (\bibinfo {year} {2004})},\ \Eprint {http://arxiv.org/abs/astro-ph/0309217}
  {arXiv:astro-ph/0309217 [astro-ph]} \BibitemShut {NoStop}%
\bibitem [{\citenamefont {{Cao}}\ \emph {et~al.}(2022)\citenamefont {{Cao}},
  \citenamefont {{Dainotti}},\ and\ \citenamefont
  {{Ratra}}}]{2022MNRAS.512..439C}%
  \BibitemOpen
  \bibfield  {author} {\bibinfo {author} {\bibfnamefont {S.}~\bibnamefont
  {{Cao}}}, \bibinfo {author} {\bibfnamefont {M.}~\bibnamefont {{Dainotti}}}, \
  and\ \bibinfo {author} {\bibfnamefont {B.}~\bibnamefont {{Ratra}}},\ }\href
  {\doibase 10.1093/mnras/stac517} {\bibfield  {journal} {\bibinfo  {journal}
  {\mnras}\ }\textbf {\bibinfo {volume} {512}},\ \bibinfo {pages} {439}
  (\bibinfo {year} {2022})},\ \Eprint {http://arxiv.org/abs/2201.05245}
  {arXiv:2201.05245 [astro-ph.CO]} \BibitemShut {NoStop}%
\bibitem [{\citenamefont {{Kumar}}\ \emph {et~al.}(2022)\citenamefont
  {{Kumar}}, \citenamefont {{Jain}}, \citenamefont {{Mahajan}}, \citenamefont
  {{Mukherjee}},\ and\ \citenamefont {{Rana}}}]{kumaretal}%
  \BibitemOpen
  \bibfield  {author} {\bibinfo {author} {\bibfnamefont {D.}~\bibnamefont
  {{Kumar}}}, \bibinfo {author} {\bibfnamefont {D.}~\bibnamefont {{Jain}}},
  \bibinfo {author} {\bibfnamefont {S.}~\bibnamefont {{Mahajan}}}, \bibinfo
  {author} {\bibfnamefont {A.}~\bibnamefont {{Mukherjee}}}, \ and\ \bibinfo
  {author} {\bibfnamefont {A.}~\bibnamefont {{Rana}}},\ }\href@noop {}
  {\bibfield  {journal} {\bibinfo  {journal} {arXiv e-prints}\ ,\ \bibinfo
  {eid} {arXiv:2205.13247}} (\bibinfo {year} {2022})},\ \Eprint
  {http://arxiv.org/abs/2205.13247} {arXiv:2205.13247 [astro-ph.CO]}
  \BibitemShut {NoStop}%
\bibitem [{\citenamefont {{Jiao}}\ \emph {et~al.}(2023)\citenamefont {{Jiao}},
  \citenamefont {{Borghi}}, \citenamefont {{Moresco}},\ and\ \citenamefont
  {{Zhang}}}]{2023ApJS..265...48J}%
  \BibitemOpen
  \bibfield  {author} {\bibinfo {author} {\bibfnamefont {K.}~\bibnamefont
  {{Jiao}}}, \bibinfo {author} {\bibfnamefont {N.}~\bibnamefont {{Borghi}}},
  \bibinfo {author} {\bibfnamefont {M.}~\bibnamefont {{Moresco}}}, \ and\
  \bibinfo {author} {\bibfnamefont {T.-J.}\ \bibnamefont {{Zhang}}},\ }\href
  {\doibase 10.3847/1538-4365/acbc77} {\bibfield  {journal} {\bibinfo
  {journal} {\apjs}\ }\textbf {\bibinfo {volume} {265}},\ \bibinfo {eid} {48}
  (\bibinfo {year} {2023})},\ \Eprint {http://arxiv.org/abs/2205.05701}
  {arXiv:2205.05701 [astro-ph.CO]} \BibitemShut {NoStop}%
\bibitem [{\citenamefont {{Salvaterra}}\ \emph {et~al.}(2009)\citenamefont
  {{Salvaterra}}, \citenamefont {{Della Valle}}, \citenamefont {{Campana}},
  \citenamefont {{Chincarini}}, \citenamefont {{Covino}} \emph
  {et~al.}}]{Salvaterra2009}%
  \BibitemOpen
  \bibfield  {author} {\bibinfo {author} {\bibfnamefont {R.}~\bibnamefont
  {{Salvaterra}}}, \bibinfo {author} {\bibfnamefont {M.}~\bibnamefont {{Della
  Valle}}}, \bibinfo {author} {\bibfnamefont {S.}~\bibnamefont {{Campana}}},
  \bibinfo {author} {\bibfnamefont {G.}~\bibnamefont {{Chincarini}}}, \bibinfo
  {author} {\bibfnamefont {S.}~\bibnamefont {{Covino}}},  \emph {et~al.},\
  }\href {\doibase 10.1038/nature08445} {\bibfield  {journal} {\bibinfo
  {journal} {\nat}\ }\textbf {\bibinfo {volume} {461}},\ \bibinfo {pages}
  {1258} (\bibinfo {year} {2009})},\ \Eprint {http://arxiv.org/abs/0906.1578}
  {arXiv:0906.1578 [astro-ph.CO]} \BibitemShut {NoStop}%
\bibitem [{\citenamefont {{Cucchiara}}\ \emph {et~al.}(2011)\citenamefont
  {{Cucchiara}}, \citenamefont {{Levan}}, \citenamefont {{Fox}}, \citenamefont
  {{Tanvir}}, \citenamefont {{Ukwatta}} \emph {et~al.}}]{Cucchiara2011}%
  \BibitemOpen
  \bibfield  {author} {\bibinfo {author} {\bibfnamefont {A.}~\bibnamefont
  {{Cucchiara}}}, \bibinfo {author} {\bibfnamefont {A.~J.}\ \bibnamefont
  {{Levan}}}, \bibinfo {author} {\bibfnamefont {D.~B.}\ \bibnamefont {{Fox}}},
  \bibinfo {author} {\bibfnamefont {N.~R.}\ \bibnamefont {{Tanvir}}}, \bibinfo
  {author} {\bibfnamefont {T.~N.}\ \bibnamefont {{Ukwatta}}},  \emph {et~al.},\
  }\href {\doibase 10.1088/0004-637X/736/1/7} {\bibfield  {journal} {\bibinfo
  {journal} {\apj}\ }\textbf {\bibinfo {volume} {736}},\ \bibinfo {eid} {7}
  (\bibinfo {year} {2011})},\ \Eprint {http://arxiv.org/abs/1105.4915}
  {arXiv:1105.4915} \BibitemShut {NoStop}%
\bibitem [{\citenamefont {{Ghirlanda}}\ \emph
  {et~al.}(2004{\natexlab{b}})\citenamefont {{Ghirlanda}}, \citenamefont
  {{Ghisellini}},\ and\ \citenamefont {{Lazzati}}}]{ghirlanda}%
  \BibitemOpen
  \bibfield  {author} {\bibinfo {author} {\bibfnamefont {G.}~\bibnamefont
  {{Ghirlanda}}}, \bibinfo {author} {\bibfnamefont {G.}~\bibnamefont
  {{Ghisellini}}}, \ and\ \bibinfo {author} {\bibfnamefont {D.}~\bibnamefont
  {{Lazzati}}},\ }\href {\doibase 10.1086/424913} {\bibfield  {journal}
  {\bibinfo  {journal} {\apj}\ }\textbf {\bibinfo {volume} {616}},\ \bibinfo
  {pages} {331} (\bibinfo {year} {2004}{\natexlab{b}})},\ \Eprint
  {http://arxiv.org/abs/astro-ph/0405602} {arXiv:astro-ph/0405602 [astro-ph]}
  \BibitemShut {NoStop}%
\bibitem [{\citenamefont {{Luongo}}\ and\ \citenamefont
  {{Muccino}}(2021{\natexlab{c}})}]{2021Galax...9...77L}%
  \BibitemOpen
  \bibfield  {author} {\bibinfo {author} {\bibfnamefont {O.}~\bibnamefont
  {{Luongo}}}\ and\ \bibinfo {author} {\bibfnamefont {M.}~\bibnamefont
  {{Muccino}}},\ }\href {\doibase 10.3390/galaxies9040077} {\bibfield
  {journal} {\bibinfo  {journal} {Galaxies}\ }\textbf {\bibinfo {volume} {9}},\
  \bibinfo {pages} {77} (\bibinfo {year} {2021}{\natexlab{c}})},\ \Eprint
  {http://arxiv.org/abs/2110.14408} {arXiv:2110.14408 [astro-ph.HE]}
  \BibitemShut {NoStop}%
\bibitem [{\citenamefont {{Band}}(2003)}]{Band2003}%
  \BibitemOpen
  \bibfield  {author} {\bibinfo {author} {\bibfnamefont {D.~L.}\ \bibnamefont
  {{Band}}},\ }\href {\doibase 10.1086/374242} {\bibfield  {journal} {\bibinfo
  {journal} {\apj}\ }\textbf {\bibinfo {volume} {588}},\ \bibinfo {pages} {945}
  (\bibinfo {year} {2003})},\ \Eprint
  {http://arxiv.org/abs/arXiv:astro-ph/0212452} {arXiv:astro-ph/0212452}
  \BibitemShut {NoStop}%
\bibitem [{\citenamefont {{Luongo}}\ and\ \citenamefont
  {{Muccino}}(2020)}]{2020A&A...641A.174L}%
  \BibitemOpen
  \bibfield  {author} {\bibinfo {author} {\bibfnamefont {O.}~\bibnamefont
  {{Luongo}}}\ and\ \bibinfo {author} {\bibfnamefont {M.}~\bibnamefont
  {{Muccino}}},\ }\href {\doibase 10.1051/0004-6361/202038264} {\bibfield
  {journal} {\bibinfo  {journal} {\aap}\ }\textbf {\bibinfo {volume} {641}},\
  \bibinfo {eid} {A174} (\bibinfo {year} {2020})},\ \Eprint
  {http://arxiv.org/abs/2010.05218} {arXiv:2010.05218 [astro-ph.CO]}
  \BibitemShut {NoStop}%
\bibitem [{\citenamefont {{Muccino}}\ \emph {et~al.}(2023)\citenamefont
  {{Muccino}}, \citenamefont {{Luongo}},\ and\ \citenamefont
  {{Jain}}}]{2023MNRAS.523.4938M}%
  \BibitemOpen
  \bibfield  {author} {\bibinfo {author} {\bibfnamefont {M.}~\bibnamefont
  {{Muccino}}}, \bibinfo {author} {\bibfnamefont {O.}~\bibnamefont {{Luongo}}},
  \ and\ \bibinfo {author} {\bibfnamefont {D.}~\bibnamefont {{Jain}}},\ }\href
  {\doibase 10.1093/mnras/stad1760} {\bibfield  {journal} {\bibinfo  {journal}
  {\mnras}\ }\textbf {\bibinfo {volume} {523}},\ \bibinfo {pages} {4938}
  (\bibinfo {year} {2023})},\ \Eprint {http://arxiv.org/abs/2208.13700}
  {arXiv:2208.13700 [astro-ph.CO]} \BibitemShut {NoStop}%
\bibitem [{\citenamefont {{Riess}}\ \emph {et~al.}(2004)\citenamefont
  {{Riess}}, \citenamefont {{Strolger}}, \citenamefont {{Tonry}}, \citenamefont
  {{Casertano}}, \citenamefont {{Ferguson}} \emph {et~al.}}]{riess2004}%
  \BibitemOpen
  \bibfield  {author} {\bibinfo {author} {\bibfnamefont {A.~G.}\ \bibnamefont
  {{Riess}}}, \bibinfo {author} {\bibfnamefont {L.-G.}\ \bibnamefont
  {{Strolger}}}, \bibinfo {author} {\bibfnamefont {J.}~\bibnamefont {{Tonry}}},
  \bibinfo {author} {\bibfnamefont {S.}~\bibnamefont {{Casertano}}}, \bibinfo
  {author} {\bibfnamefont {H.~C.}\ \bibnamefont {{Ferguson}}},  \emph
  {et~al.},\ }\href {\doibase 10.1086/383612} {\bibfield  {journal} {\bibinfo
  {journal} {\apj}\ }\textbf {\bibinfo {volume} {607}},\ \bibinfo {pages} {665}
  (\bibinfo {year} {2004})},\ \Eprint {http://arxiv.org/abs/astro-ph/0402512}
  {arXiv:astro-ph/0402512 [astro-ph]} \BibitemShut {NoStop}%
\bibitem [{\citenamefont {{Capozziello}}\ \emph {et~al.}(2019)\citenamefont
  {{Capozziello}}, \citenamefont {{D'Agostino}},\ and\ \citenamefont
  {{Luongo}}}]{capozziello2019}%
  \BibitemOpen
  \bibfield  {author} {\bibinfo {author} {\bibfnamefont {S.}~\bibnamefont
  {{Capozziello}}}, \bibinfo {author} {\bibfnamefont {R.}~\bibnamefont
  {{D'Agostino}}}, \ and\ \bibinfo {author} {\bibfnamefont {O.}~\bibnamefont
  {{Luongo}}},\ }\href {\doibase 10.1142/S0218271819300167} {\bibfield
  {journal} {\bibinfo  {journal} {International Journal of Modern Physics D}\
  }\textbf {\bibinfo {volume} {28}},\ \bibinfo {eid} {1930016} (\bibinfo {year}
  {2019})},\ \Eprint {http://arxiv.org/abs/1904.01427} {arXiv:1904.01427
  [gr-qc]} \BibitemShut {NoStop}%
\bibitem [{\citenamefont {Benetti}\ and\ \citenamefont
  {Capozziello}(2019)}]{benetti}%
  \BibitemOpen
  \bibfield  {author} {\bibinfo {author} {\bibfnamefont {M.}~\bibnamefont
  {Benetti}}\ and\ \bibinfo {author} {\bibfnamefont {S.}~\bibnamefont
  {Capozziello}},\ }\href {\doibase 10.1088/1475-7516/2019/12/008} {\bibfield
  {journal} {\bibinfo  {journal} {JCAP}\ }\textbf {\bibinfo {volume} {12}},\
  \bibinfo {pages} {008} (\bibinfo {year} {2019})},\ \Eprint
  {http://arxiv.org/abs/1910.09975} {arXiv:1910.09975 [astro-ph.CO]}
  \BibitemShut {NoStop}%
\bibitem [{\citenamefont {{Metropolis}}\ \emph {et~al.}(1953)\citenamefont
  {{Metropolis}}, \citenamefont {{Rosenbluth}}, \citenamefont {{Rosenbluth}},
  \citenamefont {{Teller}},\ and\ \citenamefont
  {{Teller}}}]{1953JChPh..21.1087M}%
  \BibitemOpen
  \bibfield  {author} {\bibinfo {author} {\bibfnamefont {N.}~\bibnamefont
  {{Metropolis}}}, \bibinfo {author} {\bibfnamefont {A.~W.}\ \bibnamefont
  {{Rosenbluth}}}, \bibinfo {author} {\bibfnamefont {M.~N.}\ \bibnamefont
  {{Rosenbluth}}}, \bibinfo {author} {\bibfnamefont {A.~H.}\ \bibnamefont
  {{Teller}}}, \ and\ \bibinfo {author} {\bibfnamefont {E.}~\bibnamefont
  {{Teller}}},\ }\href {\doibase 10.1063/1.1699114} {\bibfield  {journal}
  {\bibinfo  {journal} {\jcp}\ }\textbf {\bibinfo {volume} {21}},\ \bibinfo
  {pages} {1087} (\bibinfo {year} {1953})}\BibitemShut {NoStop}%
\bibitem [{\citenamefont {{Hastings}}(1970)}]{1970Bimka..57...97H}%
  \BibitemOpen
  \bibfield  {author} {\bibinfo {author} {\bibfnamefont {W.~K.}\ \bibnamefont
  {{Hastings}}},\ }\href {\doibase 10.1093/biomet/57.1.97} {\bibfield
  {journal} {\bibinfo  {journal} {Biometrika}\ }\textbf {\bibinfo {volume}
  {57}},\ \bibinfo {pages} {97} (\bibinfo {year} {1970})}\BibitemShut {NoStop}%
\bibitem [{\citenamefont {{Bocquet}}\ and\ \citenamefont
  {{Carter}}(2016)}]{2016JOSS....1...46B}%
  \BibitemOpen
  \bibfield  {author} {\bibinfo {author} {\bibfnamefont {S.}~\bibnamefont
  {{Bocquet}}}\ and\ \bibinfo {author} {\bibfnamefont {F.~W.}\ \bibnamefont
  {{Carter}}},\ }\href {\doibase 10.21105/joss.00046} {\bibfield  {journal}
  {\bibinfo  {journal} {The Journal of Open Source Software}\ }\textbf
  {\bibinfo {volume} {1}},\ \bibinfo {pages} {46} (\bibinfo {year}
  {2016})}\BibitemShut {NoStop}%
\bibitem [{\citenamefont {{D'Agostini}}(2005)}]{2005physics..11182D}%
  \BibitemOpen
  \bibfield  {author} {\bibinfo {author} {\bibfnamefont {G.}~\bibnamefont
  {{D'Agostini}}},\ }\href {\doibase 10.48550/arXiv.physics/0511182} {\bibfield
   {journal} {\bibinfo  {journal} {arXiv e-prints}\ ,\ \bibinfo {eid}
  {physics/0511182}} (\bibinfo {year} {2005})},\ \Eprint
  {http://arxiv.org/abs/physics/0511182} {arXiv:physics/0511182
  [physics.data-an]} \BibitemShut {NoStop}%
\bibitem [{\citenamefont {{Riess}}\ \emph {et~al.}(2018)\citenamefont
  {{Riess}}, \citenamefont {{Rodney}}, \citenamefont {{Scolnic}}, \citenamefont
  {{Shafer}}, \citenamefont {{Strolger}} \emph {et~al.}}]{2018ApJ...853..126R}%
  \BibitemOpen
  \bibfield  {author} {\bibinfo {author} {\bibfnamefont {A.~G.}\ \bibnamefont
  {{Riess}}}, \bibinfo {author} {\bibfnamefont {S.~A.}\ \bibnamefont
  {{Rodney}}}, \bibinfo {author} {\bibfnamefont {D.~M.}\ \bibnamefont
  {{Scolnic}}}, \bibinfo {author} {\bibfnamefont {D.~L.}\ \bibnamefont
  {{Shafer}}}, \bibinfo {author} {\bibfnamefont {L.-G.}\ \bibnamefont
  {{Strolger}}},  \emph {et~al.},\ }\href {\doibase 10.3847/1538-4357/aaa5a9}
  {\bibfield  {journal} {\bibinfo  {journal} {\apj}\ }\textbf {\bibinfo
  {volume} {853}},\ \bibinfo {eid} {126} (\bibinfo {year} {2018})},\ \Eprint
  {http://arxiv.org/abs/1710.00844} {arXiv:1710.00844 [astro-ph.CO]}
  \BibitemShut {NoStop}%
\bibitem [{\citenamefont {{Scolnic}}\ \emph {et~al.}(2018)\citenamefont
  {{Scolnic}}, \citenamefont {{Jones}}, \citenamefont {{Rest}}, \citenamefont
  {{Pan}}, \citenamefont {{Chornock}} \emph {et~al.}}]{2018ApJ...859..101S}%
  \BibitemOpen
  \bibfield  {author} {\bibinfo {author} {\bibfnamefont {D.~M.}\ \bibnamefont
  {{Scolnic}}}, \bibinfo {author} {\bibfnamefont {D.~O.}\ \bibnamefont
  {{Jones}}}, \bibinfo {author} {\bibfnamefont {A.}~\bibnamefont {{Rest}}},
  \bibinfo {author} {\bibfnamefont {Y.~C.}\ \bibnamefont {{Pan}}}, \bibinfo
  {author} {\bibfnamefont {R.}~\bibnamefont {{Chornock}}},  \emph {et~al.},\
  }\href {\doibase 10.3847/1538-4357/aab9bb} {\bibfield  {journal} {\bibinfo
  {journal} {\apj}\ }\textbf {\bibinfo {volume} {859}},\ \bibinfo {eid} {101}
  (\bibinfo {year} {2018})},\ \Eprint {http://arxiv.org/abs/1710.00845}
  {arXiv:1710.00845 [astro-ph.CO]} \BibitemShut {NoStop}%
\bibitem [{\citenamefont {{Planck Collaboration}}(2020)}]{planck2018}%
  \BibitemOpen
  \bibfield  {author} {\bibinfo {author} {\bibnamefont {{Planck
  Collaboration}}},\ }\href {\doibase 10.1051/0004-6361/201833910} {\bibfield
  {journal} {\bibinfo  {journal} {\aap}\ }\textbf {\bibinfo {volume} {641}},\
  \bibinfo {eid} {A6} (\bibinfo {year} {2020})},\ \Eprint
  {http://arxiv.org/abs/1807.06209} {arXiv:1807.06209 [astro-ph.CO]}
  \BibitemShut {NoStop}%
\bibitem [{\citenamefont {{Blake}}\ \emph {et~al.}(2011)\citenamefont
  {{Blake}}, \citenamefont {{Kazin}}, \citenamefont {{Beutler}}, \citenamefont
  {{Davis}}, \citenamefont {{Parkinson}} \emph {et~al.}}]{wigglez}%
  \BibitemOpen
  \bibfield  {author} {\bibinfo {author} {\bibfnamefont {C.}~\bibnamefont
  {{Blake}}}, \bibinfo {author} {\bibfnamefont {E.~A.}\ \bibnamefont
  {{Kazin}}}, \bibinfo {author} {\bibfnamefont {F.}~\bibnamefont {{Beutler}}},
  \bibinfo {author} {\bibfnamefont {T.~M.}\ \bibnamefont {{Davis}}}, \bibinfo
  {author} {\bibfnamefont {D.}~\bibnamefont {{Parkinson}}},  \emph {et~al.},\
  }\href {\doibase 10.1111/j.1365-2966.2011.19592.x} {\bibfield  {journal}
  {\bibinfo  {journal} {\mnras}\ }\textbf {\bibinfo {volume} {418}},\ \bibinfo
  {pages} {1707} (\bibinfo {year} {2011})},\ \Eprint
  {http://arxiv.org/abs/1108.2635} {arXiv:1108.2635 [astro-ph.CO]} \BibitemShut
  {NoStop}%
\bibitem [{\citenamefont {{Riess}}\ \emph {et~al.}(2022)\citenamefont
  {{Riess}}, \citenamefont {{Yuan}}, \citenamefont {{Macri}}, \citenamefont
  {{Scolnic}}, \citenamefont {{Brout}} \emph {et~al.}}]{2022ApJ...934L...7R}%
  \BibitemOpen
  \bibfield  {author} {\bibinfo {author} {\bibfnamefont {A.~G.}\ \bibnamefont
  {{Riess}}}, \bibinfo {author} {\bibfnamefont {W.}~\bibnamefont {{Yuan}}},
  \bibinfo {author} {\bibfnamefont {L.~M.}\ \bibnamefont {{Macri}}}, \bibinfo
  {author} {\bibfnamefont {D.}~\bibnamefont {{Scolnic}}}, \bibinfo {author}
  {\bibfnamefont {D.}~\bibnamefont {{Brout}}},  \emph {et~al.},\ }\href
  {\doibase 10.3847/2041-8213/ac5c5b} {\bibfield  {journal} {\bibinfo
  {journal} {\apjl}\ }\textbf {\bibinfo {volume} {934}},\ \bibinfo {eid} {L7}
  (\bibinfo {year} {2022})},\ \Eprint {http://arxiv.org/abs/2112.04510}
  {arXiv:2112.04510 [astro-ph.CO]} \BibitemShut {NoStop}%
\bibitem [{\citenamefont {{Khadka}}\ \emph {et~al.}(2021)\citenamefont
  {{Khadka}}, \citenamefont {{Luongo}}, \citenamefont {{Muccino}},\ and\
  \citenamefont {{Ratra}}}]{2021JCAP...09..042K}%
  \BibitemOpen
  \bibfield  {author} {\bibinfo {author} {\bibfnamefont {N.}~\bibnamefont
  {{Khadka}}}, \bibinfo {author} {\bibfnamefont {O.}~\bibnamefont {{Luongo}}},
  \bibinfo {author} {\bibfnamefont {M.}~\bibnamefont {{Muccino}}}, \ and\
  \bibinfo {author} {\bibfnamefont {B.}~\bibnamefont {{Ratra}}},\ }\href
  {\doibase 10.1088/1475-7516/2021/09/042} {\bibfield  {journal} {\bibinfo
  {journal} {\jcap}\ }\textbf {\bibinfo {volume} {2021}},\ \bibinfo {eid} {042}
  (\bibinfo {year} {2021})},\ \Eprint {http://arxiv.org/abs/2105.12692}
  {arXiv:2105.12692 [astro-ph.CO]} \BibitemShut {NoStop}%
\bibitem [{\citenamefont {{Muccino}}\ \emph {et~al.}(2021)\citenamefont
  {{Muccino}}, \citenamefont {{Izzo}}, \citenamefont {{Luongo}}, \citenamefont
  {{Boshkayev}}, \citenamefont {{Amati}}, \citenamefont {{Della Valle}},
  \citenamefont {{Pisani}},\ and\ \citenamefont
  {{Zaninoni}}}]{2021ApJ...908..181M}%
  \BibitemOpen
  \bibfield  {author} {\bibinfo {author} {\bibfnamefont {M.}~\bibnamefont
  {{Muccino}}}, \bibinfo {author} {\bibfnamefont {L.}~\bibnamefont {{Izzo}}},
  \bibinfo {author} {\bibfnamefont {O.}~\bibnamefont {{Luongo}}}, \bibinfo
  {author} {\bibfnamefont {K.}~\bibnamefont {{Boshkayev}}}, \bibinfo {author}
  {\bibfnamefont {L.}~\bibnamefont {{Amati}}}, \bibinfo {author} {\bibfnamefont
  {M.}~\bibnamefont {{Della Valle}}}, \bibinfo {author} {\bibfnamefont {G.~B.}\
  \bibnamefont {{Pisani}}}, \ and\ \bibinfo {author} {\bibfnamefont
  {E.}~\bibnamefont {{Zaninoni}}},\ }\href {\doibase 10.3847/1538-4357/abd254}
  {\bibfield  {journal} {\bibinfo  {journal} {\apj}\ }\textbf {\bibinfo
  {volume} {908}},\ \bibinfo {eid} {181} (\bibinfo {year} {2021})},\ \Eprint
  {http://arxiv.org/abs/2012.03392} {arXiv:2012.03392 [astro-ph.CO]}
  \BibitemShut {NoStop}%
\end{thebibliography}
\end{document}